\documentclass[amsmath,amssymb,amsbsy,reprint,prb,preprintnumbers,showpacs,superscriptaddress]{revtex4-2}
\usepackage{graphicx,color}
\usepackage{dcolumn}
\usepackage{bm}
\usepackage{braket}
\usepackage{mathtools}
\usepackage{ulem}
\usepackage[breaklinks,colorlinks=true,linkcolor=blue,urlcolor=blue,citecolor=blue]{hyperref}
\usepackage{times}
\usepackage{physics}
\usepackage{latexsym}
\usepackage{amsmath, amssymb}
\usepackage{mathtools}
\usepackage{multirow}

\newcommand{\be}{\begin{eqnarray}}
\newcommand{\ee}{\end{eqnarray}}

\renewcommand{\theequation}{\arabic{equation}}

\begin{document}
\title{
Strong and weak symmetries and their spontaneous symmetry breaking in mixed states emerging from the quantum Ising model under multiple decoherence}

%(decoherent)}
\date{\today}
\author{Takahiro Orito}%$^{2}$}
\thanks{These authors equally contributed to this work}
\affiliation{Institute for Solid State Physics, The University of Tokyo, Kashiwa, Chiba, 277-8581, Japan}
\author{Yoshihito Kuno} %$^{1}$}
\thanks{These authors equally contributed to this work}
\affiliation{Graduate School of Engineering science, Akita University, Akita 010-8502, Japan}
\author{Ikuo Ichinose} %$^{2}$}
\thanks{A professor emeritus}
\affiliation{Department of Applied Physics, Nagoya Institute of Technology, Nagoya, 466-8555, Japan}

%\affiliation{Graduate School of Engineering Science, Akita University, Akita 010-8502, Japan}

\begin{abstract} 
Discovering and categorizing quantum orders in mixed many-body systems are currently one of the most important problems. 
Specific types of decoherence applied to typical quantum many-body states can induce a novel kind of mixed state accompanying characteristic symmetry orders, which has no counterparts in pure many-body states.
We study phenomena generated by interplay between two types of decoherence applied to the one-dimensional transverse field Ising model (TFIM). 
We show that in the doubled Hilbert space formalism, the decoherence can be described by filtering operation applied to matrix product states (MPS) defined in the doubled Hilbert system. 
The filtering operation induces specific deformation of the MPS, which approximates the ground state of a certain parent Hamiltonian in the doubled Hilbert space. 
In the present case, such a parent Hamiltonian is the quantum Ashkin-Teller model, having a rich phase diagram with a critical lines and quantum phase transitions. 
By investigating the deformed MPS, we find various types of mixed states emergent from the ground states of the TFIM, and clarify phase transitions between them.
In that study, strong and weak $Z_2$ symmetries play an important role, for which we introduce efficient order parameters, such as R\'{e}nyi-2 correlators, entanglement entropy, etc., in the doubled Hilbert space.
\end{abstract}

%\pacs{67.85.Hj, 75.10.-b, 03.75.Nt}

\maketitle
%%%%%%%%%%%%%%%%%%%%%%%%%%%%%%
\section{Introduction}
Noise and decoherence \cite{gardiner2000} in quantum systems are inevitable. 
For quantum computers and quantum memories, noise and decoherence from an environment generate undesired effects and they perturb quantum states of the system~\cite{preskill2018,dennis2002,wang2003,ohno2004}. 
However, even under noises, intermediate scale quantum devices \cite{ebadi2021,bluvstein2024} are expected to exhibit great ability beyond the classical ones \cite{preskill2018}.
Surprisingly enough, such effect of noise and decoherence can lead to rich non-trivial quantum states being never created in isolated quantum systems. That is, noise and decoherence applied to pure quantum states can be an essential ingredient to produce exotic mixed quantum states, which can play an important role in quantum devices. 

Recently the generation of non-trivial mixed states having no counterpart of pure states attracts lots of interest in condensed matter physics as well as quantum information communities.  
As an example, a topologically-ordered pure state \cite{Wen_text,wen2004} tends to change to a mixed state with another type of topological order \cite{bao2023,wang2024,sohal2024,zhang2024,sang2024,Chen2024_v2,KOI2024_IMTO}.
From this point of view, behavior of symmetry protected topological (SPT) states under decoherence has been studied \cite{Lee2024,Guo-and-Ashida2024,min2024} to find that the SPT order survives in an ensemble level \cite{ma2023,ma2024}. 
In order to investigate these phenomena, we note that there are two types of symmetries; strong and weak symmetries in mixed states \cite{groot2022}. 
The notion of these symmetries can lead to some classification of mixed state orders.
Then for mixed states, we have to reexamine notion of spontaneous symmetry breaking (SSB), that is,
there are several types of SSBs in various systems, such as strong symmetry SSB, weak symmetry SSB, and 
strong to weak SSB (SWSSB), etc,~\cite{lee2023,lessa2024,sala2024,KOI2024,wang2024,sohal2024,liu2024_SSSB,guo2024,shah2024,weinstein2024,Ando2024,chen2024,stephen2024}
some of which are to be carefully defined in this work.
Getting deep understanding of relation between various SSBs and discovering and proposing concrete examples of SSB phenomena induced by decoherence are currently one of the most important problems. 
In general, exact theoretical treatment of mixed states is not easy, and some of previous studies employed Choi isomorphism technique and the doubled Hilbert state formalism~\cite{Choi1975,JAMIOLKOWSKI1972}.
By using these techniques as well as effective field theory methods, symmetry properties of certain specific mixed states are discussed~\cite{lee2023,Lee2024}. 

Following this research trend, this work clarifies some aspect of decohered states by studying specific effects of tunable multiple-type decoherence on the evolution of the ground states of the one-dimensional (1D) transverse field Ising model (TFIM). 
Interplay of the symmetries of the ground state and decoherence respecting $Z_2$ strong symmetry induces a rich mixed-state phase diagram. 
In this study, we make use of the doubled Hilbert space formalism and investigate the interplay of two kinds of decoherence: nearest-neighbor $ZZ$ and local $X$ types. 
In this doubled Hilbert space formalism, a mixed state density matrix is mapped to a state vector that is not normalized generally.
Here, we recognize decoherence applied to the mixed state vector as {\it local filtering operation}, which has been used in the analysis of pure states under perturbations, especially for matrix product states (MPS) in frustration-free models~\cite{Haegeman2015,Zhu2019}. 

Local filtering operation deforms an MPS describing
the frustration-free toric code to another MPS, which is close to the ground state of the toric code in a magnetic field derived by perturbative calculation~\cite{Castelnovo2008}. 
In this work, we suitably employ this strategy, 
that is, we first prepare the density matrix of the ground state of the 1D TFIM as an MPS in the doubled Hilbert formalism. 
For this MPS, the $ZZ$ and $X$ type multiple decoherence is applied by means of two types of local filtering operators. 
From the success of the previous works~\cite{Haegeman2015,Zhu2019}, 
we expect that the deformed MPS by the filtering is at least qualitatively close to the ground state of the quantum Ashkin-Teller (qAT) model~\cite{Kohmoto1981} derived as an effective model, even though the starting TFIM is not frustration-free. 
By the above prescription, we numerically find that the deformed MPS exhibits the SWSSB phase, corresponding to the ``partially ordered phase'' in the qAT model \cite{Kohmoto1981}. 
Moreover, by fine-tuning the parameters of the decoherence (filtering) and choosing the starting ground state of the TFIM, we numerically investigate various MPS's and phase transition between them.
To this end, the viewpoint of strong and weak symmetry, as well as R\'{e}nyi-2 correlation functions and entanglement entropy observing them, play an important role. 

The rest of this paper is organized as follows. 
In Sec.~II, we show the setting of the system in this work; 1D TFIM and two types of decoherence.
In Sec.~III, we introduce the doubled Hilbert space formalism and show the interpretation of the decoherence in this formalism.
In Sec.~VI, we perform the systematic numerical calculations by using the MPS and the filtering to the MPS for various decoherence parameters. 
Here, we find that the emergent states can be understood with the help of the ground-state phase diagram of qAT model, and we analyze the phase transitions between the deformed MPS's.
In Sec. VII, we give a summery of our numerical findings from the viewpoint of the strong and weak symmetry SSB. 
Section VIII is devoted to summery and conclusion.

\section{Set up of model and decoherence protocol}
In this work, we study effects of multiple decoherence applied to the many-body ground states of the 1D TFIM, 
Hamiltonian of which is given by
\begin{eqnarray}
H_{0}=-\sum^{L-1}_{j=0}[JZ_jZ_{j+1}+hX_j],\nonumber
\end{eqnarray}
where periodic boundary conditions are imposed and $J$, $h>0$ are parameters. 
The system possesses $Z_2$ symmetry, the generator of which is a global spin flip $\prod^{L-1}_{j=0}X_{j}$. 
At $J=h$, a phase transition takes place and the ground state is critical with the Ising CFT criticality. 
For $J/h>1$, the ground state is $Z_2$ SSB (ferromagnetic) and for $J/h<1$, a paramagnetic state emerges. 
Hereafter, we denote the ground state of $H_0$ by $|\psi_0\rangle$, and its (pure) density matrix by $\rho_0=|\psi_0\rangle\langle \psi_0|$,
and $\mathcal{H}$ is the Hilbert space of the spin-1/2 $L$-site system.

\begin{figure}[t]
\begin{center} 
\vspace{0.5cm}
\includegraphics[width=8cm]{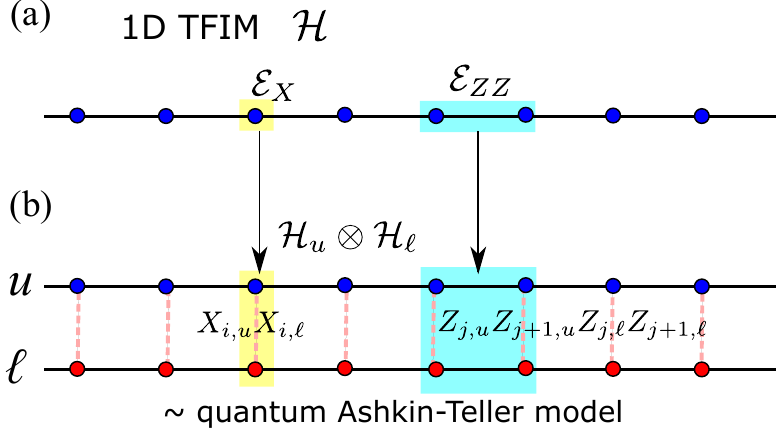}  
\end{center} 
\caption{(a) 1D quantum TFIM under two types of  decoherence. 
The state of the system is on the single Hilbert space $\mathcal{H}$. (b) The system of the 1D TFIM under decoherence is mapped to doubled Hilbert space, $\mathcal{H}_u\otimes\mathcal{H}_\ell$,
by using the Choi isomorphism, and that system is related to the quantum Ashkin-Teller model.
Multiple decoherence induces multi-body interactions in the quantum Ashkin-Teller model.}
\label{Fig1}
\end{figure}

Let us consider effects of decoherence on the ground state of the 1D TFIM. 
To this end, we introduce two types of the tunable decoherence channel applied globally to the ground state
and are given as \cite{Nielsen2011}
\begin{eqnarray}
\mathcal{E}_{ZZ}[\rho]&=&\prod^{L-1}_{j=0}\biggr[(1-p_{zz})\rho+p_{zz}Z_{j}Z_{j+1}\rho Z_{j+1}Z_{j}\biggl],\nonumber\\ 
\mathcal{E}_{X}[\rho]&=&\prod^{L-1}_{j=0}\biggr[(1-p_{x})\rho+p_{x}X_{j}\rho X_{j}\biggl], \nonumber
\end{eqnarray} 
where the strength of the decoherence is tuned by $p_{zz(x)}$, and $0\leq p_{zz(x)}\leq 1/2$. 
For $p_{zz(x)}= 1/2$, these channels correspond to projective measurements of $Z_jZ_{j+1}$ and $X_j$ without monitoring and are called maximal decoherence. 
The image of the local application of the decoherence on the 1D spin chain is shown in Fig.~\ref{Fig1} (a). Throughout this work, we study the following decohered mixed state, $\rho_D$, by the multiple decoherences 
$$
\rho_D\equiv \mathcal{E}_{ZZ}\circ \mathcal{E}_{X}[\rho_0].
$$
Note that the order in application of the above local channels is irrelevant as long as we consider decoherence channels using Pauli operators such as $\mathcal{E}_{g_j}[\rho]=(1-p)\rho+g_j \rho g^\dagger_j$, where $g_j$ is an element of Pauli group with a finite length support.
Then, general two channels are commutative, $\mathcal{E}_{g_j}\circ \mathcal{E}_{g_\ell}[\rho]=\mathcal{E}_{g_\ell}\circ \mathcal{E}_{g_j}[\rho]$ for either $[g_j,g_\ell]=0$ or $\{g_j,g_\ell\}=0$.
We investigate a mixed state, $\rho_D$, emerging through decoherence channel from the input density matrix (the ground state of the 1D TFIM) for various values of $J/h$.
In the channel, $p_{zz}$ and $p_x$ (the strength of decoherence) are parameters that determine the `phase diagram' of $\rho_D$.
%%%%%%%%%%%%%%%%%%%%%%%%%%%%%%%%%%%% 
%%%%%%%%%%%%%%%%%%%%%%%%%%%%%%%%%%%%
\section{Doubled Hilbert space formalism}
For the analysis of the decohered state $\rho_D\in \mathcal{H}$, we use the doubled Hilbert space formalism, in which the target Hilbert space is doubled as $\mathcal{H}_{u}\otimes \mathcal{H}_{\ell}$, where the subscripts $u$ and $\ell$ denote the upper and lower Hilbert spaces corresponding to ket and bra states of mixed state density matrix, respectively.
In this doubled Hilbert space formalism (Choi-Jamio\l kowski isomorphism) \cite{Choi1975,JAMIOLKOWSKI1972}, density matrix $\rho$ is vectorized, $\rho \longrightarrow |\rho\rangle\rangle$ as $|\rho\rangle\rangle\equiv \frac{1}{\sqrt{\dim[\rho]}}\sum_{k}|k\rangle\otimes \rho|k\rangle$, where $\{|k\rangle \}$ is an orthonormal set of bases in the Hilbert space $\mathcal{H}$. 
The state $|\rho\rangle\rangle$ is in the doubled Hilbert space $\mathcal{H}_u\otimes \mathcal{H}_{\ell}$.
Then, we map the density matrix $\rho_0$ of the 1D TFIM ground state to the initial state vector in the
doubled Hilbert space denoted by $|\rho_0\rangle\rangle \equiv |\psi^{*}_0\rangle|\psi_0\rangle$, where for the pure state, $|\rho_0\rangle\rangle$ is simply two copies of the ground state of the 1D TFIM $|\psi_0\rangle$, whereas the asterisk denotes the complex conjugation.

In this formalism, a general decoherence channel $\mathcal{E}$ is mapped to a (linear) operator $\hat{\mathcal{E}}$ acting on the state vector $|\rho\rangle\rangle$ in the doubled Hilbert space $\mathcal{H}_{u}\otimes \mathcal{H}_{\ell}$ ~\cite{lee2023,Lee2024} and denoted as $\hat{\mathcal{E}}|\rho\rangle\rangle$.

In general, quantum decoherence is expressed in terms of  Kraus operators; $\mathcal{E}[\rho]=\sum^{M-1}_{\alpha=0}K_{\alpha}\rho K^\dagger_{\alpha}$, where $K_\alpha$'s are Kraus operators satisfying $\sum^{M-1}_{\alpha=0}K_{\alpha}K^{\dagger}_{\alpha}=I$.
In this study, we consider the case that $K_{\alpha}$'s are Pauli operators without imaginary factor. 
In the Choi isomorphism, the channel operator is transformed as $\mathcal{E}\longrightarrow \hat{\mathcal{E}}=\sum^{M-1}_{\alpha=0}K^*_{\alpha,u}\otimes K_{\alpha,\ell}$.
Then, the two kinds of decoherence channels in the present study are given as the follows,
\begin{eqnarray}
\hat{\mathcal{E}}_{ZZ}(p_{zz})&=&\prod^{L-1}_{j=0}\biggr[(1-p_{zz})\hat{I}_{j,u}^* \otimes \hat{I}_{j,\ell}\nonumber\\ &&+p_{zz}Z_{j,u}^*Z_{j+1,u}^*\otimes Z_{j,\ell}Z_{j+1,\ell}\biggl]\nonumber\\ 
&=&\prod^{L-1}_{j=0}(1-2p_{zz})^{1/2}e^{\tau_{zz} Z_{j,u}Z_{j+1,u}\otimes Z_{j,\ell}Z_{j+1,\ell}},\nonumber\\
\hat{\mathcal{E}}_{X}(p_x)&=&\prod^{L-1}_{j=0}\biggr[(1-p_{x})\hat{I}_{j,u}^* \otimes \hat{I}_{j,\ell} +p_{x}X_{j,u}^*\otimes X_{j,\ell}\biggl]\nonumber\\
&=&\prod^{L-1}_{j=0}(1-2p_{x})^{1/2}e^{\tau_{x} X_{j,u}\otimes X_{j,\ell}},\nonumber
\end{eqnarray}
where, $\hat{I}_{j,u(\ell)}$ is an identity operator for site-$j$ vector space in $\mathcal{H}_{u(\ell)}$, $Z(X)_{j,u(\ell)}$ is Pauli-$Z$($X$) operator at site $j$ and $\tau_{zz(x)}=\tanh^{-1}[{p_{zz(x)}/(1-p_{zz(x)})}]$. 
Here, we note that the channel operator $\hat{\mathcal{E}}$ is not a unitary map in general cases although the channel is a completely-positive trace-preserving map~\cite{Nielsen2011,lee2023,Lee2024}. 
Thus, the application of the channel operator generally changes the norm of the state vector.

Please note that the initial doubled state $|\rho_0\rangle\rangle$ generated by the copy of the ground state of the 1D TFIM can be regarded as the ground state of the two decoupled TFIM on a {\it two-leg spin-1/2 ladder} shown in Fig.~\ref{Fig1}(b), where the original physical Hilbert space is doubled. 
Then the Hilbert space $\mathcal{H}_u$ describes the Hilbert space of the upper spin chain and the Hilbert space $\mathcal{H}_\ell$ describes the lower chain. 
That is, the doubled Hilbert space $\mathcal{H}_{u}\otimes \mathcal{H}_{\ell}$ corresponds to the Hilbert space of the two leg-ladder spin-1/2 system.

In the doubled Hilbert space, by using the above decoherence channel operators, the decohered state $|\rho_D\rangle\rangle$ is given by applying $\hat{\mathcal{E}}_{ZZ}(p_{zz})$ and $\hat{\mathcal{E}}_{X}(p_{x})$ to the initial state $|\rho_0\rangle\rangle$, 
\begin{eqnarray}
&&|\rho_D\rangle\rangle\equiv \hat{\mathcal{E}}_{ZZ}\hat{\mathcal{E}}_{X}|\rho_0\rangle\rangle\nonumber\\
&&= C(p_{zz},p_x,L)\prod^{L-1}_{j=0}\biggr[e^{\tau_{zz} \hat{h}^{zz}_{j,j+1}}e^{\tau_{x} \hat{h}^x_{j}}\biggl]|\rho_0\rangle\rangle.
\label{rhoD}
\end{eqnarray}
where $\hat{h}^{zz}_{j,j+1}=Z_{j,u}Z_{j+1,u}\otimes Z_{j,\ell}Z_{j+1,\ell}$, $\hat{h}^x_j=X_{j,u}\otimes X_{j,\ell}$ and $C(p_{zz},p_x,L)\equiv (1-2p_{zz})^{L/2}(1-2p_x)^{L/2}$. Note that the state $|\rho_D\rangle\rangle$ is not generally normalized, i.e., the norm $\langle\langle \rho_D|\rho_D\rangle\rangle$ corresponds to the purity $\Tr[\rho_D^2]$ ($>0$), and the norm exhibits an exponential decay with the system size $L$ due to the factor $C(p_{zz},p_x,L)$. 
This fact requires renormalization of the state vector in the calculation of some physical quantities as shown later on.

Here, we remark an important viewpoint concerning to Eq.~(\ref{rhoD}). 
Besides the factor $C(p_{zz},p_x,L)$, Eq.~(\ref{rhoD}) shows that the state $|\rho_0\rangle\rangle$ is locally-filtered by the two different kinds of local operations $e^{\tau_{zz} \hat{h}^{zz}_{j,j+1}}$ and $e^{\tau_{x} \hat{h}^x_{j}}$ and as a result, the state $|\rho_D\rangle\rangle$ emerges~\cite{Ardonne2004,CASTELNOVO2005,Castelnovo2008,Haegeman2015}. 
Filtering prescription similar to Eq.~(\ref{rhoD}) has been used to construct a state approximating a perturbed state deformed by perturbations added to a parent Hamiltonian that is typically frustration-free such as the toric code model \cite{Castelnovo2008,Haegeman2015,Zhu2019,Chen2024_v2}. 
That is, {\it in the doubled Hilbert space formalism, decoherence channel can be regarded as local filtering operation acting on density-matrix state vectors defined in the doubled Hilbert space.}
Furthermore, since the local filtering operations commute with each other, the order of their operation to the state $|\rho_0\rangle\rangle$ is irrelevant to obtain the final decohered state $|\rho_D\rangle\rangle$.

\section{Filtering operation and qualitative parent Hamiltonian}
In the previous section, we explained that the doubled-Hilbert space system can be regarded as a spin-1/2 ladder system, the Hilbert space of which is given by $\mathcal{H}_{u}\otimes \mathcal{H}_{\ell}$.
Based on this picture and the previous studies of the filtering scheme \cite{Castelnovo2008,Haegeman2015,Zhu2019,Chen2024_v2}, 
the form of the channels $\hat{\mathcal{E}}_{ZZ}$ and $\hat{\mathcal{E}}_X$ in Eq.~(\ref{rhoD}) suggests that $\hat{h}^{zz}_{j,j+1}$ and $\hat{h}^x_{j}$ can be regarded as perturbative terms to the doubled TFIM Hamiltonian of the ladder system as shown in Fig.~\ref{Fig1}(b). 
The strengths of the effective terms are proportional to $\tau_{zz}$ and $\tau_{x}$ tuned by $p_{zz}$ and $p_x$. 
Then, we expect that our target decohered state $|\rho_D\rangle\rangle$ is closely related to the ground states of 
the quantum Ashkin-Teller model \cite{Kohmoto1981}, the Hamiltonian of which is given on the ladder as follows,
\begin{eqnarray}
H_{qAT}&=&-J\sum^{L-1}_{j=0}[Z_{j,u}Z_{j+1,u}+Z_{j,\ell}Z_{j+1,\ell}\nonumber\\
&&+\lambda_{zz} Z_{j,u}Z_{j,\ell}Z_{j+1,u}Z_{j+1,\ell}]\nonumber\\
&&-h\sum^{L-1}_{j=0}[X_{j,u}+X_{j,\ell}+\lambda_x X_{j,u}X_{j,\ell}].\nonumber
\end{eqnarray}
The above Hamiltonian is derived from a highly-anisotropic version of 2D classical Ashkin-Teller model \cite{Ashkin1943,Solyom} by the time-continuum-limit formalism \cite{Kogut1979}, and then the Hamiltonian $H_{qAT}$ has $Z_2\times Z_2$ symmetry with generators $\prod X_{j,u}$ and $\prod X_{j,\ell}$.
There also exists the obvious vertical inversion symmetry between the upper and the lower chain, $u\longleftrightarrow \ell$, thus, the system is $D_4$ symmetric \cite{O'Brien2015}. Furthermore, we expect parameter relations such as $J\lambda_{zz} \longleftrightarrow \tau_{zz}(p_{zz})$ and $h\lambda_{x} \longleftrightarrow \tau_{x}(p_{x})$, which are expected to qualitatively hold.

%%%%%%%%%%%%%%%%%%%%%%%%%%%%%%%%%%%% 
\begin{figure}[t]
\begin{center} 
\vspace{0.5cm}
\includegraphics[width=6cm]{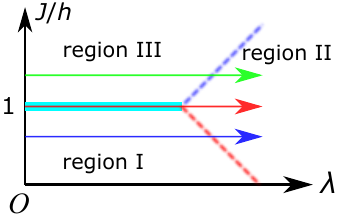}  
\end{center} 
\caption{Schematic phase diagram of the doubled system and parameter sweeps used in numerical calculations for filtering MPS. 
We expect that the global phase diagram of the doubled system is closely related to that of the quantum Ashkin-Teller model~\cite{Kohmoto1981}. 
The right blue shaded line represents a critical line separating region I (`paramagnetic phase') and region III (`ferromagnetic phase'). 
The green, red and blue solid arrows represent the target $\lambda$ parameter sweep lines with $J/h=1.2$, $1$ and $0.8$, respectively.
In our numerical calculation, this sweep of $\lambda$ is performed by parameter control of $p_{zz} (p_x)$.}
\label{Fig2}
\end{figure}
%%%%%%%%%%%%%%%%%%%%%%%%%%%%%%%%%%%%
The global ground state phase diagram of $H_{qAT}$ has been investigated in detail ~\cite{Kohmoto1981,O'Brien2015,Bridgeman2015}. 
In particular, the phase transition criticality on the line $J/h=1$ was numerically investigated in detail \cite{Bridgeman2015}. 
In the region $\lambda_{zz}=\lambda_{x}\equiv \lambda>0$ (since $\tau_{zz(x)}>0$), there are three ground-state phases \cite{Kohmoto1981,Yamanaka1994}, 
(I) Double chain spontaneous $Z_2$ symmetry broken phase (regime III), (II) Double paramagnetic phase (regime I) and (III) Diagonal $Z_2$ symmetric phase (regime II) (called ``partially-ordered phase" ~\cite{Kohmoto1981}). 
The image of the diagram is shown in Fig.~\ref{Fig2}.
In particular on the critical line between $-1/\sqrt{2}\leq \lambda \leq 1$ for $J/h=1$, the system exhibits the critical behavior described by the bosonic CFT \cite{CFT_book}.

In the previous works, this filtering method constructing perturbed states has succeeded in obtaining states close to the true ground states in various perturbed Hamiltonians~\cite{Ardonne2004,CASTELNOVO2005,Castelnovo2008,Haegeman2015}. 
Thus, we expect that the decohered state $|\rho_D\rangle\rangle$ resultant to decoherence exhibits physical properties (orders, entanglement structure, presence of phase transition, etc.) similar to those of the ground state of the qAT model $H_{qAT}$.
That is, even though the qAT model is not frustration-free, we expect that the phase diagram of the qAT model
sheds light on `phase diagram' of the decohered state $|\rho_D\rangle\rangle$ and is helpful for understanding physical properties of $|\rho_D\rangle\rangle$. 
This expectation will be verified by the numerical study given later on.

\section{Analysis of decohered state vectors in MPS formalism}
In the rest of the work, we numerically study the detailed physical properties of the decohered state $|\rho_D\rangle\rangle$ by using the MPS formalism to analyze large ladder systems and clarify entanglement properties of the decohered state vector $|\rho_D\rangle\rangle$. 
To this end, we employ the TeNPy library \cite{TeNPy,Hauschild2024}.

We first prepare an initial state $|\rho_0\rangle\rangle$ by using the DMRG searching for the ground state of the two decoupled upper and lower TFIM's on the ladder system shown in Fig.~\ref{Fig1}(b) for various values of $J$ with fixing $h=1$. 
For the obtained MPS $|\rho_0\rangle\rangle$, we apply the filtering operations $\hat{\mathcal{E}}_{ZZ}$ and $\hat{\mathcal{E}}_X$ 
to the state $|\rho_0\rangle\rangle$ as varying $p_x$ and $p_{zz}$ and obtain MPS's of $|\rho_D\rangle\rangle$.

For the practical numerical calculation, we search for the condition on the probabilities
$(p_{zz}, p_x)$ to realize the decoherence corresponding to $H_{qAT}$ with $\lambda_{zz}=\lambda_x(=\lambda)$  for various values of $J/h$. 
From the parameter correspondence discussed in the previous section, we expect $J\lambda_{zz}=c\tau_{zz}(p_{zz})$ and  $h\lambda_{x}=c\tau_{x}(p_{x})$, where $c$ is a positive constant and we set $h=1$, hereafter.
After some algebra, we find that the conditions $\lambda_{zz}=\lambda_x$ and $(1/J)\tau_{zz}(p_{zz})=\tau_x(p_x)$ are satisfied with $p_x=1/2-(1/2)(1-2p_{zz})^{1/J}$, which is an increasing function of $p_{zz}$ for $J>0$. 
In the practical protocol, we vary the value of $p_{zz}$ and fix the corresponding value of $p_x$ using the above equation, and then apply the channel operations $\hat{\mathcal{E}}_{ZZ}(p_{zz})$ and $\hat{\mathcal{E}}_X(p_x)$.
It is obvious that this procedure preserves the condition $\lambda_{zz}=\lambda_x$ in the corresponding qAT model. 
Increase of $p_{zz}$ with the relation $p_x=1/2-(1/2)(1-2p_{zz})^{1/J}$ corresponds to an increase of $\lambda$ in the qAT model.

In this work, we numerically calculate the following three observables.
The first one is the (reduced) susceptibility of R\'{e}nyi-2 correlator, 
\begin{eqnarray}
\chi^{\rm II}_{ZZ}&=&{2 \over L}\sum^{L/2}_{r=1}C^{\rm II}_{ZZ}(0,r),\nonumber
\label{chaiIIs}
\end{eqnarray}
with
\begin{eqnarray}
C^{\rm II}_{ZZ}(i,j)\equiv \frac{\langle\langle \rho_D|Z_{i,u}Z_{j,u}Z_{i,\ell}Z_{j,\ell}|\rho_D\rangle\rangle}{\langle\langle \rho_D|\rho_D\rangle\rangle},\nonumber
\end{eqnarray}
where $|\rho_D\rangle\rangle$ is an unnormalized  filtered MPS. 
In the original physical 1D system perspective, $C^{\rm II}_{ZZ}(i,j)$ corresponds to the R\'{e}nyi-2 correlator calculated with the density matrix $\rho_D$;
$$\displaystyle{C^{\rm II}_{Z_i Z_j}\equiv \frac{\Tr[Z_iZ_j\rho_D Z_jZ_i \rho_D]}{\Tr[(\rho_D)^2]}}.$$
This observable is an order parameter that detects SSB of strong symmetry but {\it not} that of the weak symmetry~\cite{lee2023,lessa2024,sala2024}.
In fact, the behavior such as, $C^{\rm II}_{ZZ}(i,j)\neq 0, C^{\rm II}_{Z_i Z_j}\neq 0$ for $|i-j| \to \infty$, 
indicates the emergence of a genuine SSB state, $\tilde{\rho}_{\rm SSB}$, with a non-vanishing one-point function 
$\Tr [Z_i\tilde{\rho}_{\rm SSB}Z_i\tilde{\rho}_{\rm SSB}]\neq 0$ for the thermodynamic limit. 
Brief explanation of strong and weak symmetries considered in this work is given in Appendix A.

%%%%%%%%%%%%%%%%%%%%%%%%%%%%%%%%%%%% 
\begin{figure*}[t]
\begin{center} 
\vspace{0.5cm}
\includegraphics[width=16cm]{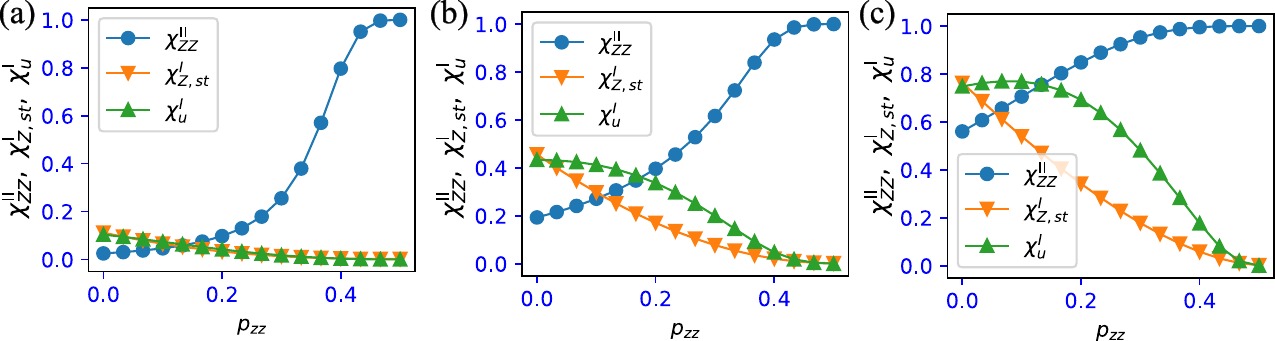}  
\end{center} 
\caption{$p_{zz}$-dependence of the sums of correlation, $\chi^{\rm II}_{ZZ}$, $\chi^{I}_{Z,st}$ and $\chi^{I}_{u}$ for $J/h=0.8$ [(a)], $1$ [(b)] and $1.2$[(c)]. Here, the value of $p_{zz}$ is related to the strength of $\lambda$ in the qAT model. The system size is $L=28$.}
\label{Fig3}
\end{figure*}
%%%%%%%%%%%%%%%%%%%%%%%%%%%%%%%%%%%%
%chis_L=28_J=0.8_1_1.2.pdf

The second observable is a correlator to characterize the $Z_2$-SSB in the doubled Hilbert space formalism, given by
$$
C^{I}_{Z,st}(i,j)=\frac{\langle\langle {\bf 1}|Z_{i,u}Z_{j,u}|\rho_D\rangle\rangle}{\langle\langle {\bf 1}|\rho_D\rangle\rangle},
$$
where $|{\bf 1}\rangle\rangle\equiv \displaystyle{\frac{1}{2^{3L/2}}\prod^{L-1}_{j=0}|t\rangle_j}$ with $|t\rangle_j=|\uparrow_u\uparrow_{\ell}\rangle_j+|\downarrow_u\downarrow_{\ell}\rangle_j$ and the corresponding quantity in the original physical Hilbert space is ${\rm Tr}[\rho_D Z_{i}Z_{j}]$.  
This relation between the above two quantities comes from the Choi isomorphism \cite{Choi1975}, 
and $C^{I}_{Z,st}(i,j)$ can be regarded as a strange correlator~\cite{Lee2024}. 
Further explanation of this point is given in Appendix B.
 
Numerically, we focus on the sum of $C^{I}_{Z,st}(i,j)$ defined by
$$
\chi^{I}_{Z,st}=\frac{2}{L}\sum^{L/2}_{r=1}C^{I}_{Z,st}(0,r).
$$
This quantity $\chi^{I}_{Z,st}$ is used as an order parameter of the weak-symmetry SSB~\cite{lessa2024}.
Then, the combination of $\chi^{\rm II}_{ZZ}$ and $\chi^{I}_{Z,st}$ can detect the SWSSB, which is recently proposed in Refs.~\cite{lee2023,lessa2024,sala2024} for strong symmetric systems~\footnote{Strictly, to define the SWSSB, we require that the initial state, target, decoherence channel, and final decohered state satisfy to be strongly-symmetric for a target on-site symmetry~\cite{lessa2024,sala2024}.}.
In the doubled Hilbert space picture, a state with $\chi^{II}_{ZZ}\sim \mathcal{O}(1)$ and $\chi^{\rm I}_{ZZ,st}\sim 0$ exhibits SSB of the off-diagonal (i.e., strong) symmetry and also the restoration of the diagonal (i.e., weak) symmetry~\cite{lee2023}.

%%%%%%%%%%%%%%%%%%%%%%%%%%%%%%%%%%%% 
\begin{figure*}[t]
\begin{center} 
\vspace{0.5cm}
\includegraphics[width=18cm]{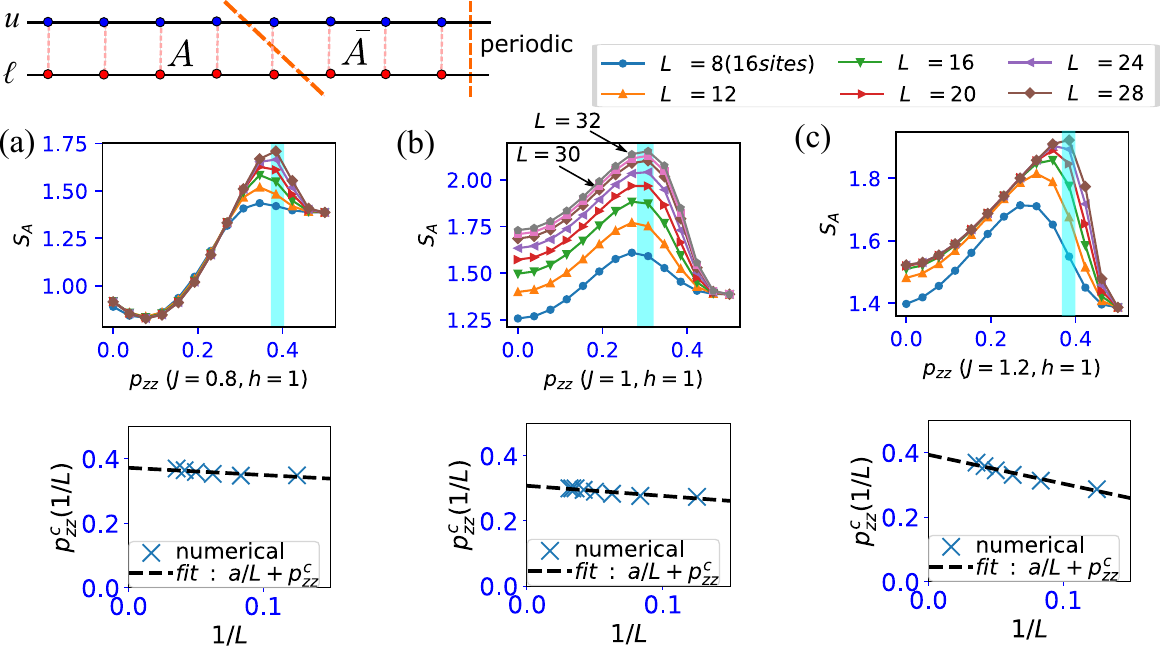}  
\end{center} 
\caption{Entanglement entropy for $J/h=0.8$ [(a)], $J/h=1$ [(b)] and $J/h=1.2$ cases [(c)]. 
Only for the case (b), we carried out the calculation up to $L=32$ system size for numerical accuracy since we start from the initial critical state.
Schematic entanglement cut for $L=8$ system under periodic boundary conditions is shown in the upper-left panel, where the orange dotted lines represent the entanglement cuts.
Panels in the bottom line show the extrapolation of data to estimate the transition probability of $p_{zz}$, denoted by $p_{zz}^c$, at which $S_A$ becomes maximum for $L\to \infty$. 
Here, to estimate locations of peaks of $S_A$, we employed a sixth-order polynomial function to estimate $p_{zz}^c$ for each system size. Using the data of finite size systems and a first-order polynomial function, $a/L+p^c_{zz}$, where $a$ and $p^c_{zz}$ are fitting parameters, we estimated the phase transition point in the thermodynamic limit.}
\label{Fig4}
\end{figure*}
%%%%%%%%%%%%%%%%%%%%%%%%%%%%%%%%%%%%
%EE_J=0.8_1_1.2_test
The third observable is the (reduced) susceptibility of the $ZZ$-correlator of the upper chain,
\begin{eqnarray}
\chi^{I}_{u}&=&{2 \over L}\sum^{L/2}_{r=1}C^{u}_{ZZ}(0,r).\nonumber
\label{chai_zzu}  
\end{eqnarray}
Here 
\begin{eqnarray}
C^{u}_{ZZ}(i,j)\equiv \frac{\langle\langle \rho_D|Z_{i,u}Z_{j,u}|\rho_D\rangle\rangle}{\langle\langle \rho_D|\rho_D\rangle\rangle}\nonumber.
\end{eqnarray}
For the original spin chain system, the above correlator $C^{u}_{ZZ}(i,j)$ corresponds to
$\Tr[\rho_DZ_iZ_j\rho_D]/\Tr[\rho_D^2]=\Tr[\rho_D^2Z_iZ_j]/\Tr[\rho_D^2]$.
Although this quantity is sightly different from the canonical correlator $\Tr[\rho_DZ_iZ_j]$,
the finite value of $\chi^{I}_{u}\sim \mathcal{O}(1)$ is expected to imply the emergence of the long-range order (LRO), i.e., $Z_2$ SSB, in the original physical 1D spin system. 
This expectation will be verified by the numerical calculation.

%%%%%%%%%%%%%%%%%%%%%%%%%%%%%%%%%%%% 
%\begin{figure}[t]
%\begin{center} 
%\vspace{0.5cm}
%\includegraphics[width=8cm]{test_fig2_test.pdf}  
%\end{center} 
%\caption{***Preliminary($J=h=1$):(a)  (b) }
%\label{Fig2}
%\end{figure}
%%%%%%%%%%%%%%%%%%%%%%%%%%%%%%%%%%%%
We also observe entanglement entropy (EE) for a subsystem (subsystem A) to study entanglement property of the decohered state $|\rho_D\rangle\rangle$, 
$$
S_A=-\Tr_{A}[\tilde{\rho}_{D,A}\ln \tilde{\rho}_{D,A}],
$$
where $\tilde{\rho}_{D,A}=\Tr_{\bar{A}}\tilde{\rho}_D$ with $\tilde{\rho}_D=|\tilde{\rho}_D\rangle\rangle \langle\langle \tilde{\rho}_D|$ where $|\tilde{\rho}_D\rangle\rangle$ is the normalized state of $|\rho_D\rangle\rangle$, $|\tilde{\rho}_D\rangle\rangle
=|\rho_D\rangle\rangle/\sqrt{\langle\langle \rho_D|\rho_D\rangle\rangle}$ \footnote{
We dare to say that EE for the doubled system is a pure mathematical object at the present time, and it is not straightforward to give some clear physical interpretation for our vertical cut EE, although the cut between the upper and lower legs of the ladder may be related to a system-environmental entanglement, in fact, the norm of the decohered state $|\rho_D\rangle\rangle$ is related to the system-environment entanglement \cite{Ashida2024}}. 
In this calculation of EE, we employ the combination of diagonal cut and vertical cut in the periodic ladder system, each two subsystems $A$ and $\bar{A}$ include $(L+1)$-sites and $(L-1)$-sits. 

A concrete example of $L=8$ ladder system is shown in the upper-left panel in Fig.~\ref{Fig4}.

%%%%%%%%%%%%%%%%%%%%%%%%%%%%%%%%%%%% 
%\begin{figure}[t]
%\begin{center} 
%\vspace{0.5cm}
%\includegraphics[width=8cm]{test_fig3_test.pdf}  
%\end{center} 
%\caption{***Preliminary($J=1.2$,$h=1$):(a)  (b) }
%\label{Fig3}
%\end{figure}
%%%%%%%%%%%%%%%%%%%%%%%%%%%%%%%%%%
\section{Numerical results by using MPS}
We investigate the decohered (filtered) state in Eq.~(\ref{rhoD}) as MPS by efficiently employing TeNPy package \cite{TeNPy,Hauschild2024}.
In Eq.~(\ref{rhoD}), the initial state $|\rho_0\rangle\rangle$, which is the ground state of the decoupled 1D TFIM in Fig.~\ref{Fig1}(b), is prepared by the DMRG. 
In all numerical simulations in this work, we set maximum bond dimension $D=200$ and truncate the singular value less than $\mathcal{O}(10^{-6})$, and the energy convergence of the iterative DMRG sweeping is $\Delta E < \mathcal{O}(10^{-4})$ to obtain the initial MPS ground state.
The filtering operation in Eq.~(\ref{rhoD}) can be also efficiently carried out by the TeNPy package. 
We then obtain the state $|\rho_D\rangle\rangle$ for each probability parameters, $p_{zz}$ and $p_x$. 
The code reliability employed in this work is examined in Appendix C.

Let us show the numerical results obtained by the protocol explained in the previous section. 
Here, we focus on three parameter sweeps of $p_{zz}$, corresponding to increasing the value of $\lambda (=\lambda_{zz}=\lambda_{x})$ from zero.
The three parameter sweep lines are (I) $J/h=0.8$, (II) $J/h=1$ (on-critical initial state), (III) $J/h=1.2$, respectively. 
The image of the parameter sweeps are shown in Fig.~\ref{Fig2}. 
The initial states for the tree sweeps (I)-(III) are a double paramagnetic state, double critical state,
and double $Z_2$-SSB state of the doubled TFIM on the ladder, respectively. 
From the parent qAT model picture, we expect that there exist three distinct regimes and ``phase transition" 
between them take place. 
These phases and the possible phase transition can be captured by the physical observables introduced in the previous section. 

We first show the behaviors of three observables $\chi^{\rm II}_{ZZ}$, $\chi^{I}_{Z,st}$ and $\chi^{I}_{u}$ for each sweep in Figs.~\ref{Fig3}(a)-~\ref{Fig3}(c). 
The data (a) ($J/h=0.8$ case) show that, for small $p_{zz}$, all values of $\chi^{\rm II}_{ZZ}$, $\chi^{I}_{Z,st}$ and $\chi^{I}_{u}$ are small reflecting the fact that we start from the trivial paramagnetic state, and therefore, that regime corresponds to the regime I of the qAT model. 
As increasing $p_{zz}$, we find that only $\chi^{\rm II}_{ZZ}$ increases implying the emergence of SWSSB phase since the weak SSB order parameter $\chi^{I}_{Z,st}$ is almost zero for large $p_{zz}$. 
We expect this phase corresponds to the regime II in the qAT phase diagram. 
Here, also the $Z_2\times Z_2$ SSB (corresponding to the independent LRO on each upper and lower chain) vanishes and $Z_2$-diagonal-symmetry restoration takes place as suggested in \cite{lee2023}.

For the data (b) ($J/h=1$ case), for small $p_{zz}$, all observables $\chi^{\rm II}_{ZZ}$, $\chi^{I}_{Z,st}$ and $\chi^{I}_{u}$ have an intermediate values. We expect the state is on critical in the qAT picture. 
As increasing $p_{zz}$, we find that only $\chi^{\rm II}_{ZZ}$ increases and, $\chi^{I}_{Z,st}$ and  $\chi^{I}_{u}$ decrease implying the emergence of the regime II, that is, SWSSB phase. 
Thus, we observe the transition from the double critical phase to the SWSSB phase.

Finally for the data (c) ($J/h=1.2$ case), for small $p_{zz}$, all observables $\chi^{\rm II}_{ZZ}$, $\chi^{I}_{Z,st}$ and $\chi^{I}_{u}$ are large. 
We expect that the state is in the regime III of the qAT phase diagram. 
From mixed state viewpoint, the state exhibits not only strong SSB with a large value of $\chi^{\rm II}_{ZZ}$ and but also  weak SSB with a large value of $\chi^{I}_{Z,st}$, indicating the emergence of the ``strong-to-trivial SSB" phase.
As increasing $p_{zz}$, we find that the state exhibits a transition into the regime II. 
Thus, we observe the phase transition from strong-to -trivial SSB phase to the SWSSB phase.

From the observation of the three quantities above, we find there are three states with distinct properties corresponding to the regimes I-III in the parent qAT model.
Then, we examine whether the changes in the three observables stem from genuine phase transitions, that is, decoherence-induced mixed state phase transitions. 
To this end, we investigate the EE for various system sizes for each of the three parameter sweeps.

The numerical results of $S_A$ are shown in Figs.~\ref{Fig4}(a)-\ref{Fig4}(c).
For the data (a) ($J/h=0.8$ case), we find that data of $S_A$ exhibit a peak around $p_{zz}\sim 0.36$
\footnote{ 
In the data in Fig.~\ref{Fig4}(a), around $p_{zz}\sim 0.1$, there is a dip of $S_A$, where there are no system-size dependence nor any signals of a phase transition. The reason for the dip of EE seems to be very complicated, and we think that it is induced by non-perturbative interplay between X+ZZ decoherences
},
and the system-size dependence of $S_A$ develops there, indicating the existence of a phase transition. 
For the locations of the observed peaks, we perform the linear extrapolation with respect to $1/L$ [See the bottom panel in Fig.~\ref{Fig4}(a)], and obtain an estimation of a transition point as $p^c_{zz}\sim 0.372$ for $L\to\infty$. 
Similarly for both the other data (b) and (c) ($J/h=1$ and $1.2$ case), $S_A$ has a peak and the value of $S_A$ at the peak increases as the system size is getting larger. 
Thus, we think that the two cases also exhibit a phase transition around $p_{zz}\sim 0.30$ for the case (b) and $p_{zz}\sim 0.39$ for the case (c), respectively. 
As the case (a), by using the linear extrapolation [See the bottom panels in Figs.~\ref{Fig4}(b) and ~\ref{Fig4}(c)], we estimate  the transition point for $L\to \infty$ as $p^c_{zz}\sim 0.308$ for $J/h=1$ case and $p^c_{zz}\sim 0.393$ for $J/h=1.2$ case, respectively.
The existence of the above phase transitions is good agreement with the phase diagram of the qAT model. 
As an additional interesting numerical result, we display the subsystem-size dependence of the EE in Appendix D.
In particular, estimation of the central charge of a possible CFT is given there.

%%%%%%%%%%%%%%%%%%%%%%%%%%%%%%%%%%
\begin{table*}[t]
\begin{tabular}{ |c||c|c|c|c| } 
\hline
    &strong SSB ($C^{\rm II}_{Z_i Z_j}$) & weak SSB ($C^{I}_{Z,st}$)& single chain LRO ($C^{u}_{ZZ}$) & \mbox{mixed state order type} \\\hline
{\rm Region I}  & $0$ & $0$ & $0$ & paramagnetic trivial \\
{\rm Region II} & $\mathcal{O}(1)$ & $0$ & $0$ & strong-to-weak SSB and diagonal $Z_2$ symmetry restoration \\
{\rm Region III} & $\mathcal{O}(1)$ & $\mathcal{O}(1)$ & $\mathcal{O}(1)$ & strong-to-trivial SSB and $Z_2 \times Z_2$ SSB \\
\hline
\end{tabular}
\caption{Summery of symmetry properties of three kinds of mixed states observed in this work. 
Here, we show the values of the correlation functions for $|i-j|\to \infty$, and ${\cal O}(1)$ denotes a finite value.}
\end{table*}

\section{Summary of mixed state order from numerical calculation of correlators}

In this section, we shall summarize the properties of the three phases obtained by numerics in the previous section. 
We first take a look at how the strong and weak $Z_2$-parity symmetries are supported in the channel and initial state.

The applied decoherence channel $\hat{\mathcal{E}}_{ZZ}\circ \hat{\mathcal{E}}_{X}$ is strong symmetric for the $Z_2$ parity symmetry [See Appendix A]. 
Thus, we expect the filtering in the doubled Hilbert space formalism respects the strong symmetry. 
Also the parent Hamiltonian (qAT model) has the $Z^u_2 \times Z^{\ell}_{2}$ symmetry indicating that the target decohered system is strong symmetric.
Then, we can discuss a possible SSB of the symmetries for various parameter regimes.

Before going into summery of pattern of the SSB, we have to carefully examine how the $Z_2$ symmetry is realized in the initial state $\rho_0$.
For $J/h>1$, the initial state $\rho_0$ obviously has a LRO for the $Z_2$ parity. 
By setting $\rho_0$ to the $Z_2$ parity $\prod X_j=+1$ cat state, the initial state $\rho_0$ can be regarded as a strong symmetric state. 
In the numerical calculation, we employed this prescription as the system is large but still finite.
On the other hand for $J/h<1$, the initial state $\rho_0$ is trivially strong symmetric under the $Z_2$ parity.

To clarify the symmetry properties of the system, we studied the filtered state in the doubled Hilbert space, and we found that the`phase diagram' has the three regimes, by observing the spin correlators and entanglement entropy.
As these phases are closely related to regimes I-III in the qAT model \cite{Kohmoto1981}, we used the same terminology for the phases that we numerically found.
As returning to the original physical Hilbert space, these filtered states correspond to three kinds of mixed states, and these mixed states can be characterized by their orders of the symmetry as summarized in Table I.

As shown in the table I, the regime I has no specific character called trivial paramagnetic mixed state. 
The region II has non-trivial properties indicated by the numerical observation, i.e., it corresponds to the $Z_2$ SWSSB phase in the mixed state picture, and in the doubled Hilbert space picture, that is the state with the diagonal $Z_2$ symmetry as observed by the correlator $C^{u}_{ZZ}$. 
The region III is also non-trivial, i.e., shown by the numerical observation, that regime corresponds to the strong-to-trivial $Z_2$ SSB phase in the mixed state picture since both strong and weak SSB order parameters are finite.
In the doubled Hilbert space picture, the state in the regime III exhibits $Z_2 \times Z_2$ SSB, as the upper and lower chains have independent LRO, which can be observed by the behavior of $C^{u}_{ZZ}$. 

It is quite interesting and also instructive to find that the ground-state phase diagram of the qAT model~\cite{Kohmoto1981} and that of the above mixed state emergent due to $X$ and $ZZ$ decoherence are quite similar.
In fact, the regions I and III correspond to the paramagnetic state without LROs and the SSB state with the LRO, respectively.
The properties of the above two regions and phases stem from the
original TFIM, and therefore, the above similarity is somewhat natural.
More interesting observation is that the region II of the SWSSB can be regarded as a counterpart of the spin-glass type phase II, 
which is generated by the sufficiently large  $\lambda$-terms in $H_{qAT}$.
In the qAT model, the $X$ and $ZZ$ operators in the $\lambda$-terms compete with each other,  and as a result, 
the spin-glass-type regime appears from the critical state, as observed by the spin-glass correlators~\cite{Kohmoto1981}.
On the the hand for the mixed states, decoherence operation is nothing but measurements without monitoring outcomes, and it generates the spin-glass type symmetry, i.e, the SWSSB.
Then, similarly to the ground state in the regime II of the qAT model, 
the glass property eliminates the ordinary LROs in the upper and lower chains, $\chi^{I}_u$ and $\chi^{I}_{Z,st}\sim 0$, 
but the glassy $Z_2$ order remains, which is captured by $\chi^{II}_{ZZ}$.
This phenomenon is expected to appear rather generally in various systems under decoherence.
In this sense, the similarity of the ground-state phase diagram of the qAT model and that of the above mixed state comes from rather specific choice of the $\lambda$-terms and multiple-decoherence protocol. This is a highlight of the present study.

\section{Summary and conclusion} 
We claimed that the decoherence is regarded as local filtering applied to MPS's in the doubled Hilbert space formalism. 
The filtering changes two decoupled states into a coupled state on the ladder spin system, the behavior of
which is close to the ground state of the qAT model. 
In certain parameter region of the multiple decoherence, the phase characterized by SWSSB appears. 
This phase emerges through the mixed state phase transition, which is close to the phase transition in the qAT model~\cite{Kohmoto1981}. 
We also expect that in $J/h<0$ case, the same SWSSB mixed phase emerges by increasing the strength of multiple decoherence, as the global phase diagram of the qAT model indicates by duality ~\cite{Yamanaka1994}.

As a future work, whether $Z_2$-orbifold boson CFT \cite{CFT_book} appears at mixed-state phase transitions is an interesting problem.
For the ground state phase transition in the qAT model, this problem has already been investigated in detail by MERA~\cite{Bridgeman2015}.
We also note that there has been growing interest in SSSB~\cite{shah2024,guo2024} within the framework of the Gorini-Kossakowski-Sudarshan-Lindblad (GKSL) equation~\cite{GKSL1,GKSL2} since weak and strong symmetries were first conceptualized in this context~\cite{Buca_2012,Albert_2014}. Thus, it is also a natural direction for future research to examine whether our findings, strong and weak SSB, could also take place in the context of the GKSL equation, taking the decohered (dissipative) Ising model~\cite{Shibata_2019} as a potential example.

Another interesting issue is to study physical meaning of the EE of the mixed state vector in the doubled Hilbert space formalism.
The present study showed that the EE is a good indicator of the phase transition.
We hope that we will report on this issue in the near future.

In this work, utility and specific character of the filtering formulation in the doubled Hilbert formalism to treat decoherence effects have been clarified.
This filtering formulation and the physical observable that we proposed in this work have broad applications to various spin models, such as XXZ chain. 
In particular, how the gapless ground state of the XXZ model  without LROs is affected by decoherence is interesting \cite{Ashida2024}. 
Since the doubled system that we consider  corresponds to a ladder spin system, the accumulated knowledge of pure-state ground state phase diagrams in various such spin systems gives us insight into existence of interesting mixed states that we do not know yet.

%\sout{In this work, utility and specific character of the multiple decoherence have been clarified.
%We expect that similar phenomena observed for the quantum Ising model will appear in other models
%under multiple decoherence, such as a 1D $Z_2$ gauge-Higgs model that attracts interest in the high-energy and condensed matter communities these days.}

%%%%%%%%%%%%%%%%%%%%%%%%%%%%%%%%
\section*{Acknowledgements}
This work is supported by JSPS KAKENHI: JP23KJ0360(T.O.) and JP23K13026(Y.K.). 
%%%%%%%%%%%%%%%%%%%%%%%%%%%%%%%%%%%%
%%%%%%%%%%%%

%%%%%%%%%%%%%%%%%%%%%%%%%%%%%%%%
\section*{Data availability}
The data and code that support the main findings of this study are available on Zenodo \cite{Zenodo}.
%%%%%%%%%%%%%%%%%%%%%%%%%%%%%%%%%%%%

%\clearpage
\renewcommand{\thesection}{A\arabic{section}} 
\renewcommand{\theequation}{A\arabic{equation}}
\renewcommand{\thefigure}{A\arabic{figure}}
\setcounter{equation}{0}
\setcounter{figure}{0}
%\section*{Appendix}
%\section*{Supplemental Material}
%\section*{\large{Supplemental Material}}

\appendix
\section*{Appendix}
\section{Strong and weak $Z_2$ symmetries}
We briefly explain two types of symmetries: strong and weak symmetries for density matrix \cite{groot2022}, especially for $Z_2$ symmetry discussed in Refs.~\cite{lee2023,lessa2024,sala2024} and focused in this work.

In general, a density matrix (mixed state) can have two distinct symmetries. 
As a concrete example, we consider $Z_2$ symmetry, the generator of which is $\{\hat{1},U_{Z_2}\}$ with $U^2_{Z_2}=\hat{1}$, and in the main text, $U_{Z_2}=\prod_{j}X_j$.
The first one is strong symmetry \cite{groot2022}
\begin{eqnarray}
U_{Z_2}\rho=e^{i\theta}\rho,\nonumber
\end{eqnarray}
where $\rho$ is a symmetric mixed state and $\theta$ is a global phase factor, $\theta\in \{0,\pi\}$. 
As the second one, weak symmetric state is defined as  
\begin{eqnarray}
U_{Z_2}\rho U^\dagger_{Z_2} =\rho.\nonumber
\end{eqnarray}
This condition is called the average or weak symmetry condition \cite{ma2024}, where the symmetry is satisfied after taking the ensemble average in the density matrix in general.

Strong and weak symmetry conditions are further defined for quantum channel. 
%Generic quantum channel is given by completely positive trace preserving (CPTP) maps. 
The operator-sum representation of the channel is given as \cite{Nielsen2011}
\begin{eqnarray}
\mathcal{E}(\rho)=\sum^{N-1}_{\ell=0}K_{\ell} \rho K^\dagger_{\ell}, \nonumber
\end{eqnarray}
where $\{K_{\ell}\}$ are a set of Kraus operators satisfying $\sum^{N-1}_{\ell=0} K^\dagger_{\ell} K_{\ell}=\hat{I}$ with $\hat{I}$ being the identity operation. 
The quantum channel $\mathcal{E}$ induces changes in mixed states. 
Here, the strong $Z_2$-symmetry condition on the channel is given as 
$$K_{\ell}U_{Z_2}=e^{i\theta} U_{Z_2} K_{\ell},$$
for any $\ell$.
On the other hand, weak symmetry condition on the channel is expressed as 
\begin{eqnarray}
U_{Z_2}\biggl[\sum_{\ell}K_{\ell} \rho K^\dagger_{\ell}\biggr]U^\dagger_{Z_2}=\mathcal{E}(\rho).\nonumber
\end{eqnarray}
This condition does not require that each Kraus operator is commutative with non-trivial generator $U_{Z_2}$.
As easily seen, a channel that satisfies the strong symmetry condition is automatically weak symmetric. 

\section{Canonical $Z_2$ correlator in the doubled Hilbert space}

For the 1D TFIM under decoherence, the canonical $Z_2$ correlator is given by ${\rm Tr}[\rho_D Z_{i}Z_{j}]$, employed as an order parameter characterizing the ordinary LRO in statistical mechanics.
As we explained in the main text, this observable detects SSB of both the strong and weak symmetries.
It is useful to have an expression corresponding to this in the doubled Hilbert space formalism to analyze mixed states.
By using the Choi isomorphism formula for density matrix $\rho$ ~\cite{Choi1975,lee2023},
$$
|\rho\rangle\rangle=\frac{1}{\sqrt{{\rm dim}[\rho]}}\sum_{k}|k\rangle \otimes \rho|k\rangle,
$$
where $\{|k\rangle\}$ is a basis set on the single $\mathcal{H}$. 
We note that the specific state $\rho=\hat{I}/2^L$ corresponds to an infinite temperature state in the physical system.
The state $\rho=\hat{I}/2^L$ can be regarded as a product state of the superposed triplet state with equal weight as shown in the main text, where we take the set of basis $\{|k\rangle\}$ as the spin z-component bases. 
Then, the canonical correlator ${\rm Tr}[\bar{\rho}_D Z_{i}Z_{j}]$ of decohered state $\bar{\rho}_D$ is expressed as  follows as simple calculation shows,
$$
{\rm Tr}[\bar{\rho}_D Z_{i}Z_{j}]={\rm Tr}[\bar{\rho}_D Z_{i}Z_{j}{\hat{I}}]=
C^{I}_{Z,st},
$$
where we have used $\langle\langle {\bf 1}|\bar{\rho}_D\rangle\rangle=1/\sqrt{2^L {\rm dim}[\bar{\rho}_D]}$. 
Thus, the correlator $C^{I}_{Z,st}$ in the doubled Hilbert system corresponds to the canonical $Z_2$ correlator,
and it gives an order parameter of the weak $Z_2$-SSB \cite{lessa2024,sala2024}.
%R2_ZZ_test_L=8_ED_vs_MPS_v2.pdf
%EE_test_L=8_ED_vs_MPS_v2.pdf

%%%%%%%%%%%%%%%%%%%%%%%%%%%%%%%%%%%% 
\begin{figure}[t]
\begin{center} 
\vspace{0.5cm}
\includegraphics[width=6cm]{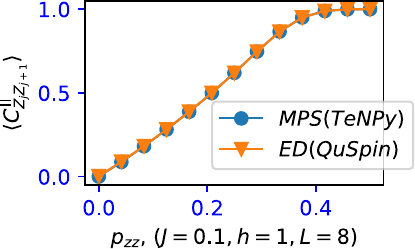}  
\end{center} 
\caption{Comparison MPS calculation with exact diagonalization by observing $\langle C^{\rm II}_{Z_j Z_{j+1}}\rangle$ as increasing the ZZ-decoherence strength $p_{zz}$. $L=8$ (total 16 sites).
Two numerical methods give the same results indicating the reliability of the present numerical methods.}
\label{FigA1}
\end{figure}
%%%%%%%%%%%%%%%%%%%%%%%%%%%%%%%%%%%%
%%%%%%%%%%%%%%%%%%%%%%%%%%%%%%%%%%%% 
\begin{figure}[t]
\begin{center} 
\vspace{0.5cm}
\includegraphics[width=6cm]{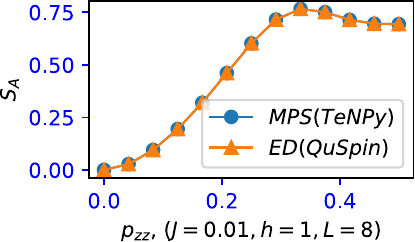}  
\end{center} 
\caption{Comparison MPS calculation with exact diagonalization by observing entanglement entropy as increasing the $ZZ$-decoherence strength $p_{zz}$. $L=8$ (total 16 sites).
Two numerical methods give the same results indicating the reliability of the present numerical methods.}
\label{FigA2}
\end{figure}
%%%%%%%%%%%%%%%%%%%%%%%%%%%%%%%%%%%%
\section{Code reliability}
We examine the reliability of our numerical technique TeNPy library \cite{TeNPy,Hauschild2024} by comparing the exact diagonalization (ED) and QuSpin package \cite{qspin1,qspin2}. 
To this end, we consider a simpler case than that of the main text. 
In the doubled Hilbert space formalism, we only consider $\hat{\mathcal{E}}_{ZZ}(p_{zz})$ decoherence and apply it to the initial doubled system $|\rho_0\rangle\rangle=|\phi^*_0\rangle|\phi_0\rangle$, where $|\phi_0\rangle$ is the unique ground state for the 1D TFIM with $J=0.1$ and $h=1$. 
%We then numerically produce $|\rho^{ZZ}_D(p_{zz})\rangle\rangle\equiv \hat{\mathcal{E}}_{ZZ}(p_{zz})|\rho_0\rangle\rangle$. 
The state $|\rho^{ZZ}_D(p_{zz})\rangle\rangle$ is obtained by both QuSpin (ED) \cite{qspin1,qspin2} and TeNPy (MPS)\cite{TeNPy,Hauschild2024}, and we observe $\langle C^{\rm II}_{Z_j Z_{j+1}}\rangle\equiv \frac{1}{L}\sum^{L-1}_{j=0}C^{\rm II}_{ZZ}(j,j+1)$. 
Figure \ref{FigA1} is the result comparing the ED and MPS for $L=8$ ladder (total $16$ sites), and 
we find the exact agreement on the results obtained by two algorithms.

As another comparison between the two methods, we calculate an entanglement entropy, where the subsystem $A$ includes only four site (one plaquette) $A=\{(0,u),(1,u),(0,\ell),(1,\ell)\}$ in the $L=8$ system. 
As a slightly different set up, we consider an initial doubled state $|\rho'_0\rangle\rangle=|\phi^{'*}_0\rangle|\phi'_0\rangle$, where $|\phi'_0\rangle$ is the unique ground state for the 1D TFIM with $J=0.01$ and $h=1$. 
The other conditions are the same with the above case. 

Figure \ref{FigA2} is the result for the entanglement entropy.
We again see an exact agreement on the results obtained by two numerical algorithms.
Thus, we conclude that our numerical methods employed in the main text are sufficiently reliable. 

\section{Subsystem size dependence of entanglement entropy for renormalized doubled state vectors}
%%%%%%%%%%%%%%%%%%%%%%%%%%%%%%%%%%%% 
\begin{figure}[t]
\begin{center} 
\vspace{0.5cm}
\includegraphics[width=7.5cm]{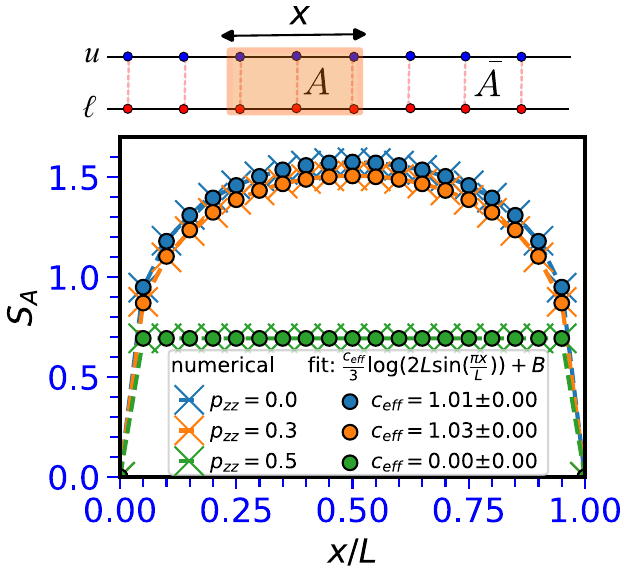}  
\end{center} 
\caption{Scaling of $S_A$  for various $p_{zz}$.
$x$ is the number of rung in subsystem (see schematic). $J=1$, $h=1$, and $L=20$ (40 sites).
}
\label{FigA3}
\end{figure}
%%%%%%%%%%%%%%%%%%%%%%%%%%%%%%%%%%%%

It is widely recognized that the scaling of $S_A$ in on-critical states obeys the logarithmic scaling \cite{CC2004}. 
Based on this, we report how the scaling of $S_A$ of a critical system is affected by the decoherence.
To examine whether the scaling of $S_A$ obeys the logarithmic scaling, we employ the following fitting function ~\cite{CC2004}, 
\begin{eqnarray}
S_A=\frac{c_{\rm eff}}{3}\log(2L\sin(\pi x/L))+B,
\label{CC}
\end{eqnarray}
where $c_{\rm eff}$ and $B$ are fitting parameters, and, $x$ is the length of the subsystem.
Here, $c_{\rm eff}$ corresponds to the effective central charge.

Figure~\ref{FigA3} shows the subsystem-size dependence of $S_A$.
Here, $x$ refers to the number of 
rung in subsystem (see schematic of Fig.~\ref{FigA3}). 
That is, we use the two vertical entanglement cuts in the periodic system~\footnote{Technically speaking, Eq.~(\ref{CC}) is employed to estimate entanglement scaling for one-dimensional systems. 
Although we numerically deal with the ladder system, the quantum state can be regarded as a one-dimensional chain system from the point of view of density matrix formalism.
From this viewpoint,
the definition of the subsystem is consistent with Eq.~(\ref{CC}).
}.
Here, we particularly focus on the critical regime, $J/h=1$.

For $p_{zz}=0$, $S_{A}(x/L)$ is well-fitted by the fitting function with $c_{\rm eff}=1$.
This result is in agreement with the scaling behavior of the critical Ising system, which obeys the logarithmic scaling with $c=1/2$~\cite{Vidal2003}. 
This result is plausible because the upper and lower critical Ising chains are totally decoupled in the case of $p_{zz}=0$, and the corresponding central charge is doubled,  that is, the sum of the effective central charges of individual chains.
For $p_{zz}=0.3$, $S_{A}(x/L)$ is also well-fitted by the fitting function with $c_{\rm eff}=1$. This result indicates that the decohered state is still in the same critical state, which is consistent with the phase diagram of the Ashkin-Teller model.
For $p_{zz}=0.5$, $S_{A}(x/L)$ shows no dependence on $x/L$ (area-law), which indicates the quantum state is no longer critical and belongs to the regime II (phase II) instead, as suggested by the observables.
Notably, given that the Ashkin-Teller model has a tricritical point, it is an interesting direction for future research to investigate whether decoherence induces comparable phase transitions.
%For $p_{zz}=0.3$, $S_{A}(x/L)$ is also well-fitted by the fitting function.
%However, the obtained $c_{\rm eff}$ slightly differs from $1$. 
%While this discrepancy may come from a finite-size effect, it could also imply the emergence of a new critical state or phase transition induced by decoherence.
%In either case, further numerical and analytical calculations will be an interesting direction for future research.  
%For $p_{zz}=0.5$, $S_{A}(x/L)$ shows no dependence on $x/L$ (area-law), which indicates the quantum state is no longer critical and belongs to the regime II (phase II) instead, as suggested by the observables.

%We observe entanglement spectrum for typical parameter points.}

\bibliography{ref}

\begin{thebibliography}{66}%
\makeatletter
\providecommand \@ifxundefined [1]{%
 \@ifx{#1\undefined}
}%
\providecommand \@ifnum [1]{%
 \ifnum #1\expandafter \@firstoftwo
 \else \expandafter \@secondoftwo
 \fi
}%
\providecommand \@ifx [1]{%
 \ifx #1\expandafter \@firstoftwo
 \else \expandafter \@secondoftwo
 \fi
}%
\providecommand \natexlab [1]{#1}%
\providecommand \enquote  [1]{``#1''}%
\providecommand \bibnamefont  [1]{#1}%
\providecommand \bibfnamefont [1]{#1}%
\providecommand \citenamefont [1]{#1}%
\providecommand \href@noop [0]{\@secondoftwo}%
\providecommand \href [0]{\begingroup \@sanitize@url \@href}%
\providecommand \@href[1]{\@@startlink{#1}\@@href}%
\providecommand \@@href[1]{\endgroup#1\@@endlink}%
\providecommand \@sanitize@url [0]{\catcode `\\12\catcode `\$12\catcode `\&12\catcode `\#12\catcode `\^12\catcode `\_12\catcode `\%12\relax}%
\providecommand \@@startlink[1]{}%
\providecommand \@@endlink[0]{}%
\providecommand \url  [0]{\begingroup\@sanitize@url \@url }%
\providecommand \@url [1]{\endgroup\@href {#1}{\urlprefix }}%
\providecommand \urlprefix  [0]{URL }%
\providecommand \Eprint [0]{\href }%
\providecommand \doibase [0]{https://doi.org/}%
\providecommand \selectlanguage [0]{\@gobble}%
\providecommand \bibinfo  [0]{\@secondoftwo}%
\providecommand \bibfield  [0]{\@secondoftwo}%
\providecommand \translation [1]{[#1]}%
\providecommand \BibitemOpen [0]{}%
\providecommand \bibitemStop [0]{}%
\providecommand \bibitemNoStop [0]{.\EOS\space}%
\providecommand \EOS [0]{\spacefactor3000\relax}%
\providecommand \BibitemShut  [1]{\csname bibitem#1\endcsname}%
\let\auto@bib@innerbib\@empty
%</preamble>
\bibitem [{\citenamefont {Gardiner}\ and\ \citenamefont {Zoller}(2000)}]{gardiner2000}%
  \BibitemOpen
  \bibfield  {author} {\bibinfo {author} {\bibfnamefont {C.~W.}\ \bibnamefont {Gardiner}}\ and\ \bibinfo {author} {\bibfnamefont {P.}~\bibnamefont {Zoller}},\ }\href@noop {} {\emph {\bibinfo {title} {Quantum Noise}}},\ \bibinfo {edition} {2nd}\ ed.,\ edited by\ \bibinfo {editor} {\bibfnamefont {H.}~\bibnamefont {Haken}}\ (\bibinfo  {publisher} {Springer},\ \bibinfo {year} {2000})\BibitemShut {NoStop}%
\bibitem [{\citenamefont {Preskill}(2018)}]{preskill2018}%
  \BibitemOpen
  \bibfield  {author} {\bibinfo {author} {\bibfnamefont {J.}~\bibnamefont {Preskill}},\ }\bibfield  {title} {\bibinfo {title} {Quantum computing in the nisq era and beyond},\ }\href {https://doi.org/10.22331/q-2018-08-06-79} {\bibfield  {journal} {\bibinfo  {journal} {Quantum}\ }\textbf {\bibinfo {volume} {2}},\ \bibinfo {pages} {79} (\bibinfo {year} {2018})}\BibitemShut {NoStop}%
\bibitem [{\citenamefont {Dennis}\ \emph {et~al.}(2002)\citenamefont {Dennis}, \citenamefont {Kitaev}, \citenamefont {Landahl},\ and\ \citenamefont {Preskill}}]{dennis2002}%
  \BibitemOpen
  \bibfield  {author} {\bibinfo {author} {\bibfnamefont {E.}~\bibnamefont {Dennis}}, \bibinfo {author} {\bibfnamefont {A.}~\bibnamefont {Kitaev}}, \bibinfo {author} {\bibfnamefont {A.}~\bibnamefont {Landahl}},\ and\ \bibinfo {author} {\bibfnamefont {J.}~\bibnamefont {Preskill}},\ }\bibfield  {title} {\bibinfo {title} {Topological quantum memory},\ }\href {https://doi.org/10.1063/1.1499754} {\bibfield  {journal} {\bibinfo  {journal} {J. Math. Phys.}\ }\textbf {\bibinfo {volume} {43}},\ \bibinfo {pages} {4452–4505} (\bibinfo {year} {2002})}\BibitemShut {NoStop}%
\bibitem [{\citenamefont {Wang}\ \emph {et~al.}(2003)\citenamefont {Wang}, \citenamefont {Harrington},\ and\ \citenamefont {Preskill}}]{wang2003}%
  \BibitemOpen
  \bibfield  {author} {\bibinfo {author} {\bibfnamefont {C.}~\bibnamefont {Wang}}, \bibinfo {author} {\bibfnamefont {J.}~\bibnamefont {Harrington}},\ and\ \bibinfo {author} {\bibfnamefont {J.}~\bibnamefont {Preskill}},\ }\bibfield  {title} {\bibinfo {title} {Confinement-higgs transition in a disordered gauge theory and the accuracy threshold for quantum memory},\ }\href {https://doi.org/10.1016/s0003-4916(02)00019-2} {\bibfield  {journal} {\bibinfo  {journal} {Ann.Phys.}\ }\textbf {\bibinfo {volume} {303}},\ \bibinfo {pages} {31–58} (\bibinfo {year} {2003})}\BibitemShut {NoStop}%
\bibitem [{\citenamefont {Ohno}\ \emph {et~al.}(2004)\citenamefont {Ohno}, \citenamefont {Arakawa}, \citenamefont {Ichinose},\ and\ \citenamefont {Matsui}}]{ohno2004}%
  \BibitemOpen
  \bibfield  {author} {\bibinfo {author} {\bibfnamefont {T.}~\bibnamefont {Ohno}}, \bibinfo {author} {\bibfnamefont {G.}~\bibnamefont {Arakawa}}, \bibinfo {author} {\bibfnamefont {I.}~\bibnamefont {Ichinose}},\ and\ \bibinfo {author} {\bibfnamefont {T.}~\bibnamefont {Matsui}},\ }\bibfield  {title} {\bibinfo {title} {Phase structure of the random-plaquette z2 gauge model: accuracy threshold for a toric quantum memory},\ }\href {https://doi.org/https://doi.org/10.1016/j.nuclphysb.2004.07.003} {\bibfield  {journal} {\bibinfo  {journal} {Nucl. Phys. B}\ }\textbf {\bibinfo {volume} {697}},\ \bibinfo {pages} {462} (\bibinfo {year} {2004})}\BibitemShut {NoStop}%
\bibitem [{\citenamefont {Ebadi}\ \emph {et~al.}(2021)\citenamefont {Ebadi}, \citenamefont {Wang}, \citenamefont {Levine}, \citenamefont {Keesling}, \citenamefont {Semeghini}, \citenamefont {Omran}, \citenamefont {Bluvstein}, \citenamefont {Samajdar}, \citenamefont {Pichler}, \citenamefont {Ho}, \citenamefont {Choi}, \citenamefont {Sachdev}, \citenamefont {Greiner}, \citenamefont {Vladan},\ and\ \citenamefont {Lukin}}]{ebadi2021}%
  \BibitemOpen
  \bibfield  {author} {\bibinfo {author} {\bibfnamefont {S.}~\bibnamefont {Ebadi}}, \bibinfo {author} {\bibfnamefont {T.~T.}\ \bibnamefont {Wang}}, \bibinfo {author} {\bibfnamefont {H.}~\bibnamefont {Levine}}, \bibinfo {author} {\bibfnamefont {A.}~\bibnamefont {Keesling}}, \bibinfo {author} {\bibfnamefont {G.}~\bibnamefont {Semeghini}}, \bibinfo {author} {\bibfnamefont {A.}~\bibnamefont {Omran}}, \bibinfo {author} {\bibfnamefont {D.}~\bibnamefont {Bluvstein}}, \bibinfo {author} {\bibfnamefont {R.}~\bibnamefont {Samajdar}}, \bibinfo {author} {\bibfnamefont {H.}~\bibnamefont {Pichler}}, \bibinfo {author} {\bibfnamefont {W.~W.}\ \bibnamefont {Ho}}, \bibinfo {author} {\bibfnamefont {S.}~\bibnamefont {Choi}}, \bibinfo {author} {\bibfnamefont {S.}~\bibnamefont {Sachdev}}, \bibinfo {author} {\bibfnamefont {M.}~\bibnamefont {Greiner}}, \bibinfo {author} {\bibfnamefont {V.}~\bibnamefont {Vladan}},\ and\ \bibinfo {author} {\bibfnamefont {M.~D.}\ \bibnamefont {Lukin}},\ }\bibfield  {title} {\bibinfo {title} {Quantum
  phases of matter on a 256-atom programmable quantum simulator},\ }\href {https://doi.org/10.1038/s41586-021-03582-4} {\bibfield  {journal} {\bibinfo  {journal} {Nature}\ }\textbf {\bibinfo {volume} {595}},\ \bibinfo {pages} {227} (\bibinfo {year} {2021})}\BibitemShut {NoStop}%
\bibitem [{\citenamefont {Bluvstein}\ \emph {et~al.}(2024)\citenamefont {Bluvstein}, \citenamefont {Evered}, \citenamefont {Geim}, \citenamefont {Li}, \citenamefont {Zhou}, \citenamefont {Manovitz}, \citenamefont {Ebadi}, \citenamefont {Cain}, \citenamefont {Kalinowski}, \citenamefont {Hangleiter}, \citenamefont {Bonilla~Ataides}, \citenamefont {Maskara}, \citenamefont {Cong}, \citenamefont {Gao}, \citenamefont {Sales~Rodriguez}, \citenamefont {Karolyshyn}, \citenamefont {Semeghini}, \citenamefont {Gullans}, \citenamefont {Greiner}, \citenamefont {Vladan},\ and\ \citenamefont {Lukin}}]{bluvstein2024}%
  \BibitemOpen
  \bibfield  {author} {\bibinfo {author} {\bibfnamefont {D.}~\bibnamefont {Bluvstein}}, \bibinfo {author} {\bibfnamefont {S.~J.}\ \bibnamefont {Evered}}, \bibinfo {author} {\bibfnamefont {A.~A.}\ \bibnamefont {Geim}}, \bibinfo {author} {\bibfnamefont {S.~H.}\ \bibnamefont {Li}}, \bibinfo {author} {\bibfnamefont {H.}~\bibnamefont {Zhou}}, \bibinfo {author} {\bibfnamefont {T.}~\bibnamefont {Manovitz}}, \bibinfo {author} {\bibfnamefont {S.}~\bibnamefont {Ebadi}}, \bibinfo {author} {\bibfnamefont {M.}~\bibnamefont {Cain}}, \bibinfo {author} {\bibfnamefont {M.}~\bibnamefont {Kalinowski}}, \bibinfo {author} {\bibfnamefont {D.}~\bibnamefont {Hangleiter}}, \bibinfo {author} {\bibfnamefont {J.~P.}\ \bibnamefont {Bonilla~Ataides}}, \bibinfo {author} {\bibfnamefont {N.}~\bibnamefont {Maskara}}, \bibinfo {author} {\bibfnamefont {I.}~\bibnamefont {Cong}}, \bibinfo {author} {\bibfnamefont {X.}~\bibnamefont {Gao}}, \bibinfo {author} {\bibfnamefont {P.}~\bibnamefont {Sales~Rodriguez}}, \bibinfo {author} {\bibfnamefont
  {T.}~\bibnamefont {Karolyshyn}}, \bibinfo {author} {\bibfnamefont {G.}~\bibnamefont {Semeghini}}, \bibinfo {author} {\bibfnamefont {M.~J.}\ \bibnamefont {Gullans}}, \bibinfo {author} {\bibfnamefont {M.}~\bibnamefont {Greiner}}, \bibinfo {author} {\bibfnamefont {V.}~\bibnamefont {Vladan}},\ and\ \bibinfo {author} {\bibfnamefont {M.~D.}\ \bibnamefont {Lukin}},\ }\bibfield  {title} {\bibinfo {title} {Logical quantum processor based on reconfigurable atom arrays},\ }\href {https://doi.org/10.1038/s41586-023-06927-3} {\bibfield  {journal} {\bibinfo  {journal} {Nature}\ }\textbf {\bibinfo {volume} {626}},\ \bibinfo {pages} {58} (\bibinfo {year} {2024})}\BibitemShut {NoStop}%
\bibitem [{\citenamefont {Zeng}\ \emph {et~al.}(2018)\citenamefont {Zeng}, \citenamefont {Chen}, \citenamefont {Zhou},\ and\ \citenamefont {Wen}}]{Wen_text}%
  \BibitemOpen
  \bibfield  {author} {\bibinfo {author} {\bibfnamefont {B.}~\bibnamefont {Zeng}}, \bibinfo {author} {\bibfnamefont {X.}~\bibnamefont {Chen}}, \bibinfo {author} {\bibfnamefont {D.-L.}\ \bibnamefont {Zhou}},\ and\ \bibinfo {author} {\bibfnamefont {X.-G.}\ \bibnamefont {Wen}},\ }\href {https://arxiv.org/abs/1508.02595} {\bibinfo {title} {Quantum information meets quantum matter -- from quantum entanglement to topological phase in many-body systems}} (\bibinfo {year} {2018}),\ \Eprint {https://arxiv.org/abs/1508.02595} {arXiv:1508.02595 [cond-mat.str-el]} \BibitemShut {NoStop}%
\bibitem [{\citenamefont {Wen}(2004)}]{wen2004}%
  \BibitemOpen
  \bibfield  {author} {\bibinfo {author} {\bibfnamefont {X.}~\bibnamefont {Wen}},\ }\href {https://books.google.co.jp/books?id=DUUrNAEACAAJ} {\bibinfo {title} {Quantum field theory of many-body systems: From the origin of sound to an origin of light and electrons}} (\bibinfo {year} {2004})\BibitemShut {NoStop}%
\bibitem [{\citenamefont {Bao}\ \emph {et~al.}(2023)\citenamefont {Bao}, \citenamefont {Fan}, \citenamefont {Vishwanath},\ and\ \citenamefont {Altman}}]{bao2023}%
  \BibitemOpen
  \bibfield  {author} {\bibinfo {author} {\bibfnamefont {Y.}~\bibnamefont {Bao}}, \bibinfo {author} {\bibfnamefont {R.}~\bibnamefont {Fan}}, \bibinfo {author} {\bibfnamefont {A.}~\bibnamefont {Vishwanath}},\ and\ \bibinfo {author} {\bibfnamefont {E.}~\bibnamefont {Altman}},\ }\href {https://arxiv.org/abs/2301.05687} {\bibinfo {title} {Mixed-state topological order and the errorfield double formulation of decoherence-induced transitions}} (\bibinfo {year} {2023}),\ \Eprint {https://arxiv.org/abs/2301.05687} {arXiv:2301.05687 [quant-ph]} \BibitemShut {NoStop}%
\bibitem [{\citenamefont {Wang}\ \emph {et~al.}(2024)\citenamefont {Wang}, \citenamefont {Wu},\ and\ \citenamefont {Wang}}]{wang2024}%
  \BibitemOpen
  \bibfield  {author} {\bibinfo {author} {\bibfnamefont {Z.}~\bibnamefont {Wang}}, \bibinfo {author} {\bibfnamefont {Z.}~\bibnamefont {Wu}},\ and\ \bibinfo {author} {\bibfnamefont {Z.}~\bibnamefont {Wang}},\ }\href {https://arxiv.org/abs/2307.13758} {\bibinfo {title} {Intrinsic mixed-state quantum topological order}} (\bibinfo {year} {2024}),\ \Eprint {https://arxiv.org/abs/2307.13758} {arXiv:2307.13758 [quant-ph]} \BibitemShut {NoStop}%
\bibitem [{\citenamefont {Sohal}\ and\ \citenamefont {Prem}(2024)}]{sohal2024}%
  \BibitemOpen
  \bibfield  {author} {\bibinfo {author} {\bibfnamefont {R.}~\bibnamefont {Sohal}}\ and\ \bibinfo {author} {\bibfnamefont {A.}~\bibnamefont {Prem}},\ }\href {https://arxiv.org/abs/2403.13879} {\bibinfo {title} {A noisy approach to intrinsically mixed-state topological order}} (\bibinfo {year} {2024}),\ \Eprint {https://arxiv.org/abs/2403.13879} {arXiv:2403.13879 [cond-mat.str-el]} \BibitemShut {NoStop}%
\bibitem [{\citenamefont {Zhang}\ \emph {et~al.}(2024)\citenamefont {Zhang}, \citenamefont {Xu}, \citenamefont {Zhang}, \citenamefont {Xu}, \citenamefont {Bi},\ and\ \citenamefont {Luo}}]{zhang2024}%
  \BibitemOpen
  \bibfield  {author} {\bibinfo {author} {\bibfnamefont {C.}~\bibnamefont {Zhang}}, \bibinfo {author} {\bibfnamefont {Y.}~\bibnamefont {Xu}}, \bibinfo {author} {\bibfnamefont {J.-H.}\ \bibnamefont {Zhang}}, \bibinfo {author} {\bibfnamefont {C.}~\bibnamefont {Xu}}, \bibinfo {author} {\bibfnamefont {Z.}~\bibnamefont {Bi}},\ and\ \bibinfo {author} {\bibfnamefont {Z.-X.}\ \bibnamefont {Luo}},\ }\href {https://arxiv.org/abs/2409.17530} {\bibinfo {title} {Strong-to-weak spontaneous breaking of 1-form symmetry and intrinsically mixed topological order}} (\bibinfo {year} {2024}),\ \Eprint {https://arxiv.org/abs/2409.17530} {arXiv:2409.17530 [quant-ph]} \BibitemShut {NoStop}%
\bibitem [{\citenamefont {Sang}\ \emph {et~al.}(2024)\citenamefont {Sang}, \citenamefont {Zou},\ and\ \citenamefont {Hsieh}}]{sang2024}%
  \BibitemOpen
  \bibfield  {author} {\bibinfo {author} {\bibfnamefont {S.}~\bibnamefont {Sang}}, \bibinfo {author} {\bibfnamefont {Y.}~\bibnamefont {Zou}},\ and\ \bibinfo {author} {\bibfnamefont {T.~H.}\ \bibnamefont {Hsieh}},\ }\bibfield  {title} {\bibinfo {title} {Mixed-state quantum phases: Renormalization and quantum error correction},\ }\href {https://doi.org/10.1103/PhysRevX.14.031044} {\bibfield  {journal} {\bibinfo  {journal} {Phys. Rev. X}\ }\textbf {\bibinfo {volume} {14}},\ \bibinfo {pages} {031044} (\bibinfo {year} {2024})}\BibitemShut {NoStop}%
\bibitem [{\citenamefont {Chen}\ and\ \citenamefont {Grover}(2024)}]{Chen2024_v2}%
  \BibitemOpen
  \bibfield  {author} {\bibinfo {author} {\bibfnamefont {Y.-H.}\ \bibnamefont {Chen}}\ and\ \bibinfo {author} {\bibfnamefont {T.}~\bibnamefont {Grover}},\ }\bibfield  {title} {\bibinfo {title} {Unconventional topological mixed-state transition and critical phase induced by self-dual coherent errors},\ }\href {https://doi.org/10.1103/PhysRevB.110.125152} {\bibfield  {journal} {\bibinfo  {journal} {Phys. Rev. B}\ }\textbf {\bibinfo {volume} {110}},\ \bibinfo {pages} {125152} (\bibinfo {year} {2024})}\BibitemShut {NoStop}%
\bibitem [{\citenamefont {Kuno}\ \emph {et~al.}(2024{\natexlab{a}})\citenamefont {Kuno}, \citenamefont {Orito},\ and\ \citenamefont {Ichinose}}]{KOI2024_IMTO}%
  \BibitemOpen
  \bibfield  {author} {\bibinfo {author} {\bibfnamefont {Y.}~\bibnamefont {Kuno}}, \bibinfo {author} {\bibfnamefont {T.}~\bibnamefont {Orito}},\ and\ \bibinfo {author} {\bibfnamefont {I.}~\bibnamefont {Ichinose}},\ }\href {https://arxiv.org/abs/2410.14258} {\bibinfo {title} {Intrinsic mixed state topological order in a stabilizer system under stochastic decoherence}} (\bibinfo {year} {2024}{\natexlab{a}}),\ \Eprint {https://arxiv.org/abs/2410.14258} {arXiv:2410.14258 [quant-ph]} \BibitemShut {NoStop}%
\bibitem [{\citenamefont {Lee}\ \emph {et~al.}(2024)\citenamefont {Lee}, \citenamefont {You},\ and\ \citenamefont {Xu}}]{Lee2024}%
  \BibitemOpen
  \bibfield  {author} {\bibinfo {author} {\bibfnamefont {J.~Y.}\ \bibnamefont {Lee}}, \bibinfo {author} {\bibfnamefont {Y.-Z.}\ \bibnamefont {You}},\ and\ \bibinfo {author} {\bibfnamefont {C.}~\bibnamefont {Xu}},\ }\href {https://arxiv.org/abs/2210.16323} {\bibinfo {title} {Symmetry protected topological phases under decoherence}} (\bibinfo {year} {2024}),\ \Eprint {https://arxiv.org/abs/2210.16323} {arXiv:2210.16323 [cond-mat.str-el]} \BibitemShut {NoStop}%
\bibitem [{\citenamefont {Guo}\ and\ \citenamefont {Ashida}(2024)}]{Guo-and-Ashida2024}%
  \BibitemOpen
  \bibfield  {author} {\bibinfo {author} {\bibfnamefont {Y.}~\bibnamefont {Guo}}\ and\ \bibinfo {author} {\bibfnamefont {Y.}~\bibnamefont {Ashida}},\ }\bibfield  {title} {\bibinfo {title} {Two-dimensional symmetry-protected topological phases and transitions in open quantum systems},\ }\href {https://doi.org/10.1103/PhysRevB.109.195420} {\bibfield  {journal} {\bibinfo  {journal} {Phys. Rev. B}\ }\textbf {\bibinfo {volume} {109}},\ \bibinfo {pages} {195420} (\bibinfo {year} {2024})}\BibitemShut {NoStop}%
\bibitem [{\citenamefont {Min}\ \emph {et~al.}(2024)\citenamefont {Min}, \citenamefont {Zhang}, \citenamefont {Guo}, \citenamefont {Segal},\ and\ \citenamefont {Ashida}}]{min2024}%
  \BibitemOpen
  \bibfield  {author} {\bibinfo {author} {\bibfnamefont {B.}~\bibnamefont {Min}}, \bibinfo {author} {\bibfnamefont {Y.}~\bibnamefont {Zhang}}, \bibinfo {author} {\bibfnamefont {Y.}~\bibnamefont {Guo}}, \bibinfo {author} {\bibfnamefont {D.}~\bibnamefont {Segal}},\ and\ \bibinfo {author} {\bibfnamefont {Y.}~\bibnamefont {Ashida}},\ }\href {https://arxiv.org/abs/2412.02733} {\bibinfo {title} {Mixed-state phase transitions in spin-holstein models}} (\bibinfo {year} {2024}),\ \Eprint {https://arxiv.org/abs/2412.02733} {arXiv:2412.02733 [cond-mat.str-el]} \BibitemShut {NoStop}%
\bibitem [{\citenamefont {Ma}\ and\ \citenamefont {Wang}(2023)}]{ma2023}%
  \BibitemOpen
  \bibfield  {author} {\bibinfo {author} {\bibfnamefont {R.}~\bibnamefont {Ma}}\ and\ \bibinfo {author} {\bibfnamefont {C.}~\bibnamefont {Wang}},\ }\bibfield  {title} {\bibinfo {title} {Average symmetry-protected topological phases},\ }\href {https://doi.org/10.1103/PhysRevX.13.031016} {\bibfield  {journal} {\bibinfo  {journal} {Phys. Rev. X}\ }\textbf {\bibinfo {volume} {13}},\ \bibinfo {pages} {031016} (\bibinfo {year} {2023})}\BibitemShut {NoStop}%
\bibitem [{\citenamefont {Ma}\ \emph {et~al.}(2024)\citenamefont {Ma}, \citenamefont {Zhang}, \citenamefont {Bi}, \citenamefont {Cheng},\ and\ \citenamefont {Wang}}]{ma2024}%
  \BibitemOpen
  \bibfield  {author} {\bibinfo {author} {\bibfnamefont {R.}~\bibnamefont {Ma}}, \bibinfo {author} {\bibfnamefont {J.-H.}\ \bibnamefont {Zhang}}, \bibinfo {author} {\bibfnamefont {Z.}~\bibnamefont {Bi}}, \bibinfo {author} {\bibfnamefont {M.}~\bibnamefont {Cheng}},\ and\ \bibinfo {author} {\bibfnamefont {C.}~\bibnamefont {Wang}},\ }\href {https://arxiv.org/abs/2305.16399} {\bibinfo {title} {Topological phases with average symmetries: the decohered, the disordered, and the intrinsic}} (\bibinfo {year} {2024}),\ \Eprint {https://arxiv.org/abs/2305.16399} {arXiv:2305.16399 [cond-mat.str-el]} \BibitemShut {NoStop}%
\bibitem [{\citenamefont {de~Groot}\ \emph {et~al.}(2022)\citenamefont {de~Groot}, \citenamefont {Turzillo},\ and\ \citenamefont {Schuch}}]{groot2022}%
  \BibitemOpen
  \bibfield  {author} {\bibinfo {author} {\bibfnamefont {C.}~\bibnamefont {de~Groot}}, \bibinfo {author} {\bibfnamefont {A.}~\bibnamefont {Turzillo}},\ and\ \bibinfo {author} {\bibfnamefont {N.}~\bibnamefont {Schuch}},\ }\bibfield  {title} {\bibinfo {title} {Symmetry protected topological order in open quantum systems},\ }\href {https://doi.org/10.22331/q-2022-11-10-856} {\bibfield  {journal} {\bibinfo  {journal} {Quantum}\ }\textbf {\bibinfo {volume} {6}},\ \bibinfo {pages} {856} (\bibinfo {year} {2022})}\BibitemShut {NoStop}%
\bibitem [{\citenamefont {Lee}\ \emph {et~al.}(2023)\citenamefont {Lee}, \citenamefont {Jian},\ and\ \citenamefont {Xu}}]{lee2023}%
  \BibitemOpen
  \bibfield  {author} {\bibinfo {author} {\bibfnamefont {J.~Y.}\ \bibnamefont {Lee}}, \bibinfo {author} {\bibfnamefont {C.-M.}\ \bibnamefont {Jian}},\ and\ \bibinfo {author} {\bibfnamefont {C.}~\bibnamefont {Xu}},\ }\bibfield  {title} {\bibinfo {title} {Quantum criticality under decoherence or weak measurement},\ }\href {https://doi.org/10.1103/PRXQuantum.4.030317} {\bibfield  {journal} {\bibinfo  {journal} {Phys. Rev. X Quantum}\ }\textbf {\bibinfo {volume} {4}},\ \bibinfo {pages} {030317} (\bibinfo {year} {2023})}\BibitemShut {NoStop}%
\bibitem [{\citenamefont {Lessa}\ \emph {et~al.}(2024)\citenamefont {Lessa}, \citenamefont {Ma}, \citenamefont {Zhang}, \citenamefont {Bi}, \citenamefont {Cheng},\ and\ \citenamefont {Wang}}]{lessa2024}%
  \BibitemOpen
  \bibfield  {author} {\bibinfo {author} {\bibfnamefont {L.~A.}\ \bibnamefont {Lessa}}, \bibinfo {author} {\bibfnamefont {R.}~\bibnamefont {Ma}}, \bibinfo {author} {\bibfnamefont {J.-H.}\ \bibnamefont {Zhang}}, \bibinfo {author} {\bibfnamefont {Z.}~\bibnamefont {Bi}}, \bibinfo {author} {\bibfnamefont {M.}~\bibnamefont {Cheng}},\ and\ \bibinfo {author} {\bibfnamefont {C.}~\bibnamefont {Wang}},\ }\href {https://arxiv.org/abs/2405.03639} {\bibinfo {title} {Strong-to-weak spontaneous symmetry breaking in mixed quantum states}} (\bibinfo {year} {2024}),\ \Eprint {https://arxiv.org/abs/2405.03639} {arXiv:2405.03639 [quant-ph]} \BibitemShut {NoStop}%
\bibitem [{\citenamefont {Sala}\ \emph {et~al.}(2024)\citenamefont {Sala}, \citenamefont {Gopalakrishnan}, \citenamefont {Oshikawa},\ and\ \citenamefont {You}}]{sala2024}%
  \BibitemOpen
  \bibfield  {author} {\bibinfo {author} {\bibfnamefont {P.}~\bibnamefont {Sala}}, \bibinfo {author} {\bibfnamefont {S.}~\bibnamefont {Gopalakrishnan}}, \bibinfo {author} {\bibfnamefont {M.}~\bibnamefont {Oshikawa}},\ and\ \bibinfo {author} {\bibfnamefont {Y.}~\bibnamefont {You}},\ }\bibfield  {title} {\bibinfo {title} {Spontaneous strong symmetry breaking in open systems: Purification perspective},\ }\href {https://doi.org/10.1103/PhysRevB.110.155150} {\bibfield  {journal} {\bibinfo  {journal} {Phys. Rev. B}\ }\textbf {\bibinfo {volume} {110}},\ \bibinfo {pages} {155150} (\bibinfo {year} {2024})}\BibitemShut {NoStop}%
\bibitem [{\citenamefont {Kuno}\ \emph {et~al.}(2024{\natexlab{b}})\citenamefont {Kuno}, \citenamefont {Orito},\ and\ \citenamefont {Ichinose}}]{KOI2024}%
  \BibitemOpen
  \bibfield  {author} {\bibinfo {author} {\bibfnamefont {Y.}~\bibnamefont {Kuno}}, \bibinfo {author} {\bibfnamefont {T.}~\bibnamefont {Orito}},\ and\ \bibinfo {author} {\bibfnamefont {I.}~\bibnamefont {Ichinose}},\ }\bibfield  {title} {\bibinfo {title} {Strong-to-weak symmetry breaking states in stochastic dephasing stabilizer circuits},\ }\href {https://doi.org/10.1103/PhysRevB.110.094106} {\bibfield  {journal} {\bibinfo  {journal} {Phys. Rev. B}\ }\textbf {\bibinfo {volume} {110}},\ \bibinfo {pages} {094106} (\bibinfo {year} {2024}{\natexlab{b}})}\BibitemShut {NoStop}%
\bibitem [{\citenamefont {Liu}\ \emph {et~al.}(2024)\citenamefont {Liu}, \citenamefont {Chen}, \citenamefont {Zhang}, \citenamefont {Zhou},\ and\ \citenamefont {Zhang}}]{liu2024_SSSB}%
  \BibitemOpen
  \bibfield  {author} {\bibinfo {author} {\bibfnamefont {Z.}~\bibnamefont {Liu}}, \bibinfo {author} {\bibfnamefont {L.}~\bibnamefont {Chen}}, \bibinfo {author} {\bibfnamefont {Y.}~\bibnamefont {Zhang}}, \bibinfo {author} {\bibfnamefont {S.}~\bibnamefont {Zhou}},\ and\ \bibinfo {author} {\bibfnamefont {P.}~\bibnamefont {Zhang}},\ }\href {https://arxiv.org/abs/2410.09327} {\bibinfo {title} {Diagnosing strong-to-weak symmetry breaking via wightman correlators}} (\bibinfo {year} {2024}),\ \Eprint {https://arxiv.org/abs/2410.09327} {arXiv:2410.09327 [quant-ph]} \BibitemShut {NoStop}%
\bibitem [{\citenamefont {Guo}\ and\ \citenamefont {Yang}(2024)}]{guo2024}%
  \BibitemOpen
  \bibfield  {author} {\bibinfo {author} {\bibfnamefont {Y.}~\bibnamefont {Guo}}\ and\ \bibinfo {author} {\bibfnamefont {S.}~\bibnamefont {Yang}},\ }\href {https://arxiv.org/abs/2410.13734} {\bibinfo {title} {Strong-to-weak spontaneous symmetry breaking meets average symmetry-protected topological order}} (\bibinfo {year} {2024}),\ \Eprint {https://arxiv.org/abs/2410.13734} {arXiv:2410.13734 [cond-mat.str-el]} \BibitemShut {NoStop}%
\bibitem [{\citenamefont {Shah}\ \emph {et~al.}(2024)\citenamefont {Shah}, \citenamefont {Fechisin}, \citenamefont {Wang}, \citenamefont {Iosue}, \citenamefont {Watson}, \citenamefont {Wang}, \citenamefont {Ware}, \citenamefont {Gorshkov},\ and\ \citenamefont {Lin}}]{shah2024}%
  \BibitemOpen
  \bibfield  {author} {\bibinfo {author} {\bibfnamefont {J.}~\bibnamefont {Shah}}, \bibinfo {author} {\bibfnamefont {C.}~\bibnamefont {Fechisin}}, \bibinfo {author} {\bibfnamefont {Y.-X.}\ \bibnamefont {Wang}}, \bibinfo {author} {\bibfnamefont {J.~T.}\ \bibnamefont {Iosue}}, \bibinfo {author} {\bibfnamefont {J.~D.}\ \bibnamefont {Watson}}, \bibinfo {author} {\bibfnamefont {Y.-Q.}\ \bibnamefont {Wang}}, \bibinfo {author} {\bibfnamefont {B.}~\bibnamefont {Ware}}, \bibinfo {author} {\bibfnamefont {A.~V.}\ \bibnamefont {Gorshkov}},\ and\ \bibinfo {author} {\bibfnamefont {C.-J.}\ \bibnamefont {Lin}},\ }\href {https://arxiv.org/abs/2410.12900} {\bibinfo {title} {Instability of steady-state mixed-state symmetry-protected topological order to strong-to-weak spontaneous symmetry breaking}} (\bibinfo {year} {2024}),\ \Eprint {https://arxiv.org/abs/2410.12900} {arXiv:2410.12900 [quant-ph]} \BibitemShut {NoStop}%
\bibitem [{\citenamefont {Weinstein}(2024)}]{weinstein2024}%
  \BibitemOpen
  \bibfield  {author} {\bibinfo {author} {\bibfnamefont {Z.}~\bibnamefont {Weinstein}},\ }\href {https://arxiv.org/abs/2410.23512} {\bibinfo {title} {Efficient detection of strong-to-weak spontaneous symmetry breaking via the r\'enyi-1 correlator}} (\bibinfo {year} {2024}),\ \Eprint {https://arxiv.org/abs/2410.23512} {arXiv:2410.23512 [quant-ph]} \BibitemShut {NoStop}%
\bibitem [{\citenamefont {Ando}\ \emph {et~al.}(2024)\citenamefont {Ando}, \citenamefont {Ryu},\ and\ \citenamefont {Watanabe}}]{Ando2024}%
  \BibitemOpen
  \bibfield  {author} {\bibinfo {author} {\bibfnamefont {T.}~\bibnamefont {Ando}}, \bibinfo {author} {\bibfnamefont {S.}~\bibnamefont {Ryu}},\ and\ \bibinfo {author} {\bibfnamefont {M.}~\bibnamefont {Watanabe}},\ }\href {https://arxiv.org/abs/2411.04360} {\bibinfo {title} {Gauge theory and mixed state criticality}} (\bibinfo {year} {2024}),\ \Eprint {https://arxiv.org/abs/2411.04360} {arXiv:2411.04360 [cond-mat.str-el]} \BibitemShut {NoStop}%
\bibitem [{\citenamefont {Chen}\ \emph {et~al.}(2024)\citenamefont {Chen}, \citenamefont {Sun},\ and\ \citenamefont {Zhang}}]{chen2024}%
  \BibitemOpen
  \bibfield  {author} {\bibinfo {author} {\bibfnamefont {L.}~\bibnamefont {Chen}}, \bibinfo {author} {\bibfnamefont {N.}~\bibnamefont {Sun}},\ and\ \bibinfo {author} {\bibfnamefont {P.}~\bibnamefont {Zhang}},\ }\href {https://arxiv.org/abs/2411.05364} {\bibinfo {title} {Strong-to-weak symmetry breaking and entanglement transitions}} (\bibinfo {year} {2024}),\ \Eprint {https://arxiv.org/abs/2411.05364} {arXiv:2411.05364 [quant-ph]} \BibitemShut {NoStop}%
\bibitem [{\citenamefont {Stephen}\ \emph {et~al.}(2024)\citenamefont {Stephen}, \citenamefont {Nandkishore},\ and\ \citenamefont {Zhang}}]{stephen2024}%
  \BibitemOpen
  \bibfield  {author} {\bibinfo {author} {\bibfnamefont {D.~T.}\ \bibnamefont {Stephen}}, \bibinfo {author} {\bibfnamefont {R.}~\bibnamefont {Nandkishore}},\ and\ \bibinfo {author} {\bibfnamefont {J.-H.}\ \bibnamefont {Zhang}},\ }\href {https://arxiv.org/abs/2410.23354} {\bibinfo {title} {Many-body quantum catalysts for transforming between phases of matter}} (\bibinfo {year} {2024}),\ \Eprint {https://arxiv.org/abs/2410.23354} {arXiv:2410.23354 [quant-ph]} \BibitemShut {NoStop}%
\bibitem [{\citenamefont {Choi}(1975)}]{Choi1975}%
  \BibitemOpen
  \bibfield  {author} {\bibinfo {author} {\bibfnamefont {M.-D.}\ \bibnamefont {Choi}},\ }\bibfield  {title} {\bibinfo {title} {Completely positive linear maps on complex matrices},\ }\href {https://doi.org/https://doi.org/10.1016/0024-3795(75)90075-0} {\bibfield  {journal} {\bibinfo  {journal} {Lin. Alg. Appl.}\ }\textbf {\bibinfo {volume} {10}},\ \bibinfo {pages} {285} (\bibinfo {year} {1975})}\BibitemShut {NoStop}%
\bibitem [{\citenamefont {Jamiołkowski}(1972)}]{JAMIOLKOWSKI1972}%
  \BibitemOpen
  \bibfield  {author} {\bibinfo {author} {\bibfnamefont {A.}~\bibnamefont {Jamiołkowski}},\ }\bibfield  {title} {\bibinfo {title} {Linear transformations which preserve trace and positive semidefiniteness of operators},\ }\href {https://doi.org/https://doi.org/10.1016/0034-4877(72)90011-0} {\bibfield  {journal} {\bibinfo  {journal} {Rep. Math. Phys.}\ }\textbf {\bibinfo {volume} {3}},\ \bibinfo {pages} {275} (\bibinfo {year} {1972})}\BibitemShut {NoStop}%
\bibitem [{\citenamefont {Haegeman}\ \emph {et~al.}(2015)\citenamefont {Haegeman}, \citenamefont {Van~Acoleyen}, \citenamefont {Schuch}, \citenamefont {Cirac},\ and\ \citenamefont {Verstraete}}]{Haegeman2015}%
  \BibitemOpen
  \bibfield  {author} {\bibinfo {author} {\bibfnamefont {J.}~\bibnamefont {Haegeman}}, \bibinfo {author} {\bibfnamefont {K.}~\bibnamefont {Van~Acoleyen}}, \bibinfo {author} {\bibfnamefont {N.}~\bibnamefont {Schuch}}, \bibinfo {author} {\bibfnamefont {J.~I.}\ \bibnamefont {Cirac}},\ and\ \bibinfo {author} {\bibfnamefont {F.}~\bibnamefont {Verstraete}},\ }\bibfield  {title} {\bibinfo {title} {Gauging quantum states: From global to local symmetries in many-body systems},\ }\href {https://doi.org/10.1103/PhysRevX.5.011024} {\bibfield  {journal} {\bibinfo  {journal} {Phys. Rev. X}\ }\textbf {\bibinfo {volume} {5}},\ \bibinfo {pages} {011024} (\bibinfo {year} {2015})}\BibitemShut {NoStop}%
\bibitem [{\citenamefont {Zhu}\ and\ \citenamefont {Zhang}(2019)}]{Zhu2019}%
  \BibitemOpen
  \bibfield  {author} {\bibinfo {author} {\bibfnamefont {G.-Y.}\ \bibnamefont {Zhu}}\ and\ \bibinfo {author} {\bibfnamefont {G.-M.}\ \bibnamefont {Zhang}},\ }\bibfield  {title} {\bibinfo {title} {Gapless coulomb state emerging from a self-dual topological tensor-network state},\ }\href {https://doi.org/10.1103/PhysRevLett.122.176401} {\bibfield  {journal} {\bibinfo  {journal} {Phys. Rev. Lett.}\ }\textbf {\bibinfo {volume} {122}},\ \bibinfo {pages} {176401} (\bibinfo {year} {2019})}\BibitemShut {NoStop}%
\bibitem [{\citenamefont {Castelnovo}\ and\ \citenamefont {Chamon}(2008)}]{Castelnovo2008}%
  \BibitemOpen
  \bibfield  {author} {\bibinfo {author} {\bibfnamefont {C.}~\bibnamefont {Castelnovo}}\ and\ \bibinfo {author} {\bibfnamefont {C.}~\bibnamefont {Chamon}},\ }\bibfield  {title} {\bibinfo {title} {Quantum topological phase transition at the microscopic level},\ }\href {https://doi.org/10.1103/PhysRevB.77.054433} {\bibfield  {journal} {\bibinfo  {journal} {Phys. Rev. B}\ }\textbf {\bibinfo {volume} {77}},\ \bibinfo {pages} {054433} (\bibinfo {year} {2008})}\BibitemShut {NoStop}%
\bibitem [{\citenamefont {Kohmoto}\ \emph {et~al.}(1981)\citenamefont {Kohmoto}, \citenamefont {den Nijs},\ and\ \citenamefont {Kadanoff}}]{Kohmoto1981}%
  \BibitemOpen
  \bibfield  {author} {\bibinfo {author} {\bibfnamefont {M.}~\bibnamefont {Kohmoto}}, \bibinfo {author} {\bibfnamefont {M.}~\bibnamefont {den Nijs}},\ and\ \bibinfo {author} {\bibfnamefont {L.~P.}\ \bibnamefont {Kadanoff}},\ }\bibfield  {title} {\bibinfo {title} {Hamiltonian studies of the $d=2$ ashkin-teller model},\ }\href {https://doi.org/10.1103/PhysRevB.24.5229} {\bibfield  {journal} {\bibinfo  {journal} {Phys. Rev. B}\ }\textbf {\bibinfo {volume} {24}},\ \bibinfo {pages} {5229} (\bibinfo {year} {1981})}\BibitemShut {NoStop}%
\bibitem [{\citenamefont {Nielsen}\ and\ \citenamefont {Chuang}(2011)}]{Nielsen2011}%
  \BibitemOpen
  \bibfield  {author} {\bibinfo {author} {\bibfnamefont {M.~A.}\ \bibnamefont {Nielsen}}\ and\ \bibinfo {author} {\bibfnamefont {I.~L.}\ \bibnamefont {Chuang}},\ }\href@noop {} {\emph {\bibinfo {title} {Quantum Computation and Quantum Information}}},\ \bibinfo {edition} {10th}\ ed.\ (\bibinfo  {publisher} {Cambridge University Press},\ \bibinfo {address} {USA},\ \bibinfo {year} {2011})\BibitemShut {NoStop}%
\bibitem [{\citenamefont {Ardonne}\ \emph {et~al.}(2004)\citenamefont {Ardonne}, \citenamefont {Fendley},\ and\ \citenamefont {Fradkin}}]{Ardonne2004}%
  \BibitemOpen
  \bibfield  {author} {\bibinfo {author} {\bibfnamefont {E.}~\bibnamefont {Ardonne}}, \bibinfo {author} {\bibfnamefont {P.}~\bibnamefont {Fendley}},\ and\ \bibinfo {author} {\bibfnamefont {E.}~\bibnamefont {Fradkin}},\ }\bibfield  {title} {\bibinfo {title} {Topological order and conformal quantum critical points},\ }\href {https://doi.org/https://doi.org/10.1016/j.aop.2004.01.004} {\bibfield  {journal} {\bibinfo  {journal} {Ann.Phys.}\ }\textbf {\bibinfo {volume} {310}},\ \bibinfo {pages} {493} (\bibinfo {year} {2004})}\BibitemShut {NoStop}%
\bibitem [{\citenamefont {Castelnovo}\ \emph {et~al.}(2005)\citenamefont {Castelnovo}, \citenamefont {Chamon}, \citenamefont {Mudry},\ and\ \citenamefont {Pujol}}]{CASTELNOVO2005}%
  \BibitemOpen
  \bibfield  {author} {\bibinfo {author} {\bibfnamefont {C.}~\bibnamefont {Castelnovo}}, \bibinfo {author} {\bibfnamefont {C.}~\bibnamefont {Chamon}}, \bibinfo {author} {\bibfnamefont {C.}~\bibnamefont {Mudry}},\ and\ \bibinfo {author} {\bibfnamefont {P.}~\bibnamefont {Pujol}},\ }\bibfield  {title} {\bibinfo {title} {From quantum mechanics to classical statistical physics: Generalized rokhsar–kivelson hamiltonians and the “stochastic matrix form” decomposition},\ }\href {https://doi.org/10.1016/j.aop.2005.01.006} {\bibfield  {journal} {\bibinfo  {journal} {Ann.Phys.}\ }\textbf {\bibinfo {volume} {318}},\ \bibinfo {pages} {316–344} (\bibinfo {year} {2005})}\BibitemShut {NoStop}%
\bibitem [{\citenamefont {Ashkin}\ and\ \citenamefont {Teller}(1943)}]{Ashkin1943}%
  \BibitemOpen
  \bibfield  {author} {\bibinfo {author} {\bibfnamefont {J.}~\bibnamefont {Ashkin}}\ and\ \bibinfo {author} {\bibfnamefont {E.}~\bibnamefont {Teller}},\ }\bibfield  {title} {\bibinfo {title} {Statistics of two-dimensional lattices with four components},\ }\href {https://doi.org/10.1103/PhysRev.64.178} {\bibfield  {journal} {\bibinfo  {journal} {Phys. Rev.}\ }\textbf {\bibinfo {volume} {64}},\ \bibinfo {pages} {178} (\bibinfo {year} {1943})}\BibitemShut {NoStop}%
\bibitem [{\citenamefont {S\'olyom}(1981)}]{Solyom}%
  \BibitemOpen
  \bibfield  {author} {\bibinfo {author} {\bibfnamefont {J.}~\bibnamefont {S\'olyom}},\ }\bibfield  {title} {\bibinfo {title} {Duality of the block transformation and decimation for quantum spin systems},\ }\href {https://doi.org/10.1103/PhysRevB.24.230} {\bibfield  {journal} {\bibinfo  {journal} {Phys. Rev. B}\ }\textbf {\bibinfo {volume} {24}},\ \bibinfo {pages} {230} (\bibinfo {year} {1981})}\BibitemShut {NoStop}%
\bibitem [{\citenamefont {Kogut}(1979)}]{Kogut1979}%
  \BibitemOpen
  \bibfield  {author} {\bibinfo {author} {\bibfnamefont {J.~B.}\ \bibnamefont {Kogut}},\ }\bibfield  {title} {\bibinfo {title} {An introduction to lattice gauge theory and spin systems},\ }\href {https://doi.org/10.1103/RevModPhys.51.659} {\bibfield  {journal} {\bibinfo  {journal} {Rev. Mod. Phys.}\ }\textbf {\bibinfo {volume} {51}},\ \bibinfo {pages} {659} (\bibinfo {year} {1979})}\BibitemShut {NoStop}%
\bibitem [{\citenamefont {O'Brien}\ \emph {et~al.}(2015)\citenamefont {O'Brien}, \citenamefont {Bartlett}, \citenamefont {Doherty},\ and\ \citenamefont {Flammia}}]{O'Brien2015}%
  \BibitemOpen
  \bibfield  {author} {\bibinfo {author} {\bibfnamefont {A.}~\bibnamefont {O'Brien}}, \bibinfo {author} {\bibfnamefont {S.~D.}\ \bibnamefont {Bartlett}}, \bibinfo {author} {\bibfnamefont {A.~C.}\ \bibnamefont {Doherty}},\ and\ \bibinfo {author} {\bibfnamefont {S.~T.}\ \bibnamefont {Flammia}},\ }\bibfield  {title} {\bibinfo {title} {Symmetry-respecting real-space renormalization for the quantum ashkin-teller model},\ }\href {https://doi.org/10.1103/PhysRevE.92.042163} {\bibfield  {journal} {\bibinfo  {journal} {Phys. Rev. E}\ }\textbf {\bibinfo {volume} {92}},\ \bibinfo {pages} {042163} (\bibinfo {year} {2015})}\BibitemShut {NoStop}%
\bibitem [{\citenamefont {Bridgeman}\ \emph {et~al.}(2015)\citenamefont {Bridgeman}, \citenamefont {O'Brien}, \citenamefont {Bartlett},\ and\ \citenamefont {Doherty}}]{Bridgeman2015}%
  \BibitemOpen
  \bibfield  {author} {\bibinfo {author} {\bibfnamefont {J.~C.}\ \bibnamefont {Bridgeman}}, \bibinfo {author} {\bibfnamefont {A.}~\bibnamefont {O'Brien}}, \bibinfo {author} {\bibfnamefont {S.~D.}\ \bibnamefont {Bartlett}},\ and\ \bibinfo {author} {\bibfnamefont {A.~C.}\ \bibnamefont {Doherty}},\ }\bibfield  {title} {\bibinfo {title} {Multiscale entanglement renormalization ansatz for spin chains with continuously varying criticality},\ }\href {https://doi.org/10.1103/PhysRevB.91.165129} {\bibfield  {journal} {\bibinfo  {journal} {Phys. Rev. B}\ }\textbf {\bibinfo {volume} {91}},\ \bibinfo {pages} {165129} (\bibinfo {year} {2015})}\BibitemShut {NoStop}%
\bibitem [{\citenamefont {Yamanaka}\ \emph {et~al.}(1994)\citenamefont {Yamanaka}, \citenamefont {Hatsugai},\ and\ \citenamefont {Kohmoto}}]{Yamanaka1994}%
  \BibitemOpen
  \bibfield  {author} {\bibinfo {author} {\bibfnamefont {M.}~\bibnamefont {Yamanaka}}, \bibinfo {author} {\bibfnamefont {Y.}~\bibnamefont {Hatsugai}},\ and\ \bibinfo {author} {\bibfnamefont {M.}~\bibnamefont {Kohmoto}},\ }\bibfield  {title} {\bibinfo {title} {Phase diagram of the ashkin-teller quantum spin chain},\ }\href {https://doi.org/10.1103/PhysRevB.50.559} {\bibfield  {journal} {\bibinfo  {journal} {Phys. Rev. B}\ }\textbf {\bibinfo {volume} {50}},\ \bibinfo {pages} {559} (\bibinfo {year} {1994})}\BibitemShut {NoStop}%
\bibitem [{\citenamefont {{Di Francesco}}\ \emph {et~al.}(1997)\citenamefont {{Di Francesco}}, \citenamefont {Mathieu},\ and\ \citenamefont {S{\'e}n{\'e}chal}}]{CFT_book}%
  \BibitemOpen
  \bibfield  {author} {\bibinfo {author} {\bibfnamefont {P.}~\bibnamefont {{Di Francesco}}}, \bibinfo {author} {\bibfnamefont {P.}~\bibnamefont {Mathieu}},\ and\ \bibinfo {author} {\bibfnamefont {D.}~\bibnamefont {S{\'e}n{\'e}chal}},\ }\href {https://doi.org/10.1007/978-1-4612-2256-9} {\emph {\bibinfo {title} {Conformal field theory}}},\ Graduate Texts in Contemporary Physics\ (\bibinfo  {publisher} {Springer},\ \bibinfo {address} {Germany},\ \bibinfo {year} {1997})\BibitemShut {NoStop}%
\bibitem [{\citenamefont {Hauschild}\ and\ \citenamefont {Pollmann}(2018)}]{TeNPy}%
  \BibitemOpen
  \bibfield  {author} {\bibinfo {author} {\bibfnamefont {J.}~\bibnamefont {Hauschild}}\ and\ \bibinfo {author} {\bibfnamefont {F.}~\bibnamefont {Pollmann}},\ }\bibfield  {title} {\bibinfo {title} {{Efficient numerical simulations with Tensor Networks: Tensor Network Python (TeNPy)}},\ }\href {https://doi.org/10.21468/SciPostPhysLectNotes.5} {\bibfield  {journal} {\bibinfo  {journal} {SciPost Phys. Lect. Notes}\ ,\ \bibinfo {pages} {5}} (\bibinfo {year} {2018})}\BibitemShut {NoStop}%
\bibitem [{\citenamefont {Hauschild}\ \emph {et~al.}(2024)\citenamefont {Hauschild}, \citenamefont {Unfried}, \citenamefont {Anand}, \citenamefont {Andrews}, \citenamefont {Bintz}, \citenamefont {Borla}, \citenamefont {Divic}, \citenamefont {Drescher}, \citenamefont {Geiger}, \citenamefont {Hefel}, \citenamefont {Hémery}, \citenamefont {Kadow}, \citenamefont {Kemp}, \citenamefont {Kirchner}, \citenamefont {Liu}, \citenamefont {Möller}, \citenamefont {Parker}, \citenamefont {Rader}, \citenamefont {Romen}, \citenamefont {Scalet}, \citenamefont {Schoonderwoerd}, \citenamefont {Schulz}, \citenamefont {Soejima}, \citenamefont {Thoma}, \citenamefont {Wu}, \citenamefont {Zechmann}, \citenamefont {Zweng}, \citenamefont {Mong}, \citenamefont {Zaletel},\ and\ \citenamefont {Pollmann}}]{Hauschild2024}%
  \BibitemOpen
  \bibfield  {author} {\bibinfo {author} {\bibfnamefont {J.}~\bibnamefont {Hauschild}}, \bibinfo {author} {\bibfnamefont {J.}~\bibnamefont {Unfried}}, \bibinfo {author} {\bibfnamefont {S.}~\bibnamefont {Anand}}, \bibinfo {author} {\bibfnamefont {B.}~\bibnamefont {Andrews}}, \bibinfo {author} {\bibfnamefont {M.}~\bibnamefont {Bintz}}, \bibinfo {author} {\bibfnamefont {U.}~\bibnamefont {Borla}}, \bibinfo {author} {\bibfnamefont {S.}~\bibnamefont {Divic}}, \bibinfo {author} {\bibfnamefont {M.}~\bibnamefont {Drescher}}, \bibinfo {author} {\bibfnamefont {J.}~\bibnamefont {Geiger}}, \bibinfo {author} {\bibfnamefont {M.}~\bibnamefont {Hefel}}, \bibinfo {author} {\bibfnamefont {K.}~\bibnamefont {Hémery}}, \bibinfo {author} {\bibfnamefont {W.}~\bibnamefont {Kadow}}, \bibinfo {author} {\bibfnamefont {J.}~\bibnamefont {Kemp}}, \bibinfo {author} {\bibfnamefont {N.}~\bibnamefont {Kirchner}}, \bibinfo {author} {\bibfnamefont {V.~S.}\ \bibnamefont {Liu}}, \bibinfo {author} {\bibfnamefont {G.}~\bibnamefont {Möller}},
  \bibinfo {author} {\bibfnamefont {D.}~\bibnamefont {Parker}}, \bibinfo {author} {\bibfnamefont {M.}~\bibnamefont {Rader}}, \bibinfo {author} {\bibfnamefont {A.}~\bibnamefont {Romen}}, \bibinfo {author} {\bibfnamefont {S.}~\bibnamefont {Scalet}}, \bibinfo {author} {\bibfnamefont {L.}~\bibnamefont {Schoonderwoerd}}, \bibinfo {author} {\bibfnamefont {M.}~\bibnamefont {Schulz}}, \bibinfo {author} {\bibfnamefont {T.}~\bibnamefont {Soejima}}, \bibinfo {author} {\bibfnamefont {P.}~\bibnamefont {Thoma}}, \bibinfo {author} {\bibfnamefont {Y.}~\bibnamefont {Wu}}, \bibinfo {author} {\bibfnamefont {P.}~\bibnamefont {Zechmann}}, \bibinfo {author} {\bibfnamefont {L.}~\bibnamefont {Zweng}}, \bibinfo {author} {\bibfnamefont {R.~S.~K.}\ \bibnamefont {Mong}}, \bibinfo {author} {\bibfnamefont {M.~P.}\ \bibnamefont {Zaletel}},\ and\ \bibinfo {author} {\bibfnamefont {F.}~\bibnamefont {Pollmann}},\ }\href {https://arxiv.org/abs/2408.02010} {\bibinfo {title} {Tensor network python (tenpy) version 1}} (\bibinfo {year} {2024}),\
  \Eprint {https://arxiv.org/abs/2408.02010} {arXiv:2408.02010 [cond-mat.str-el]} \BibitemShut {NoStop}%
\bibitem [{Note1()}]{Note1}%
  \BibitemOpen
  \bibinfo {note} {Strictly, to define the SWSSB, we require that the initial state, target, decoherence channel, and final decohered state satisfy to be strongly-symmetric for a target on-site symmetry~\cite {lessa2024,sala2024}.}\BibitemShut {Stop}%
\bibitem [{Note2()}]{Note2}%
  \BibitemOpen
  \bibinfo {note} {We dare to say that EE for the doubled system is a pure mathematical object at the present time, and it is not straightforward to give some clear physical interpretation for our vertical cut EE, although the cut between the upper and lower legs of the ladder may be related to a system-environmental entanglement, in fact, the norm of the decohered state $|\rho _D\rangle \rangle $ is related to the system-environment entanglement \cite {Ashida2024}}\BibitemShut {NoStop}%
\bibitem [{Note3()}]{Note3}%
  \BibitemOpen
  \bibinfo {note} {In the data in Fig.~\ref {Fig4}(a), around $p_{zz}\sim 0.1$, there is a dip of $S_A$, where there are no system-size dependence nor any signals of a phase transition. The reason for the dip of EE seems to be very complicated, and we think that it is induced by non-perturbative interplay between X+ZZ decoherences}\BibitemShut {NoStop}%
\bibitem [{\citenamefont {Lindblad}(1976)}]{GKSL1}%
  \BibitemOpen
  \bibfield  {author} {\bibinfo {author} {\bibfnamefont {G.}~\bibnamefont {Lindblad}},\ }\bibfield  {title} {\bibinfo {title} {On the generators of quantum dynamical semigroups},\ }\href {https://doi.org/10.1007/BF01608499} {\bibfield  {journal} {\bibinfo  {journal} {Commun. Math. Phys.}\ }\textbf {\bibinfo {volume} {48}},\ \bibinfo {pages} {119} (\bibinfo {year} {1976})}\BibitemShut {NoStop}%
\bibitem [{\citenamefont {Gorini}\ \emph {et~al.}(1976)\citenamefont {Gorini}, \citenamefont {Kossakowski},\ and\ \citenamefont {Sudarshan}}]{GKSL2}%
  \BibitemOpen
  \bibfield  {author} {\bibinfo {author} {\bibfnamefont {V.}~\bibnamefont {Gorini}}, \bibinfo {author} {\bibfnamefont {A.}~\bibnamefont {Kossakowski}},\ and\ \bibinfo {author} {\bibfnamefont {E.~C.~G.}\ \bibnamefont {Sudarshan}},\ }\bibfield  {title} {\bibinfo {title} {Completely positive dynamical semigroups of n‐level systems},\ }\href {https://doi.org/10.1063/1.522979} {\bibfield  {journal} {\bibinfo  {journal} {J. Math. Phys.}\ }\textbf {\bibinfo {volume} {17}},\ \bibinfo {pages} {821} (\bibinfo {year} {1976})}\BibitemShut {NoStop}%
\bibitem [{\citenamefont {Buča}\ and\ \citenamefont {Prosen}(2012)}]{Buca_2012}%
  \BibitemOpen
  \bibfield  {author} {\bibinfo {author} {\bibfnamefont {B.}~\bibnamefont {Buča}}\ and\ \bibinfo {author} {\bibfnamefont {T.}~\bibnamefont {Prosen}},\ }\bibfield  {title} {\bibinfo {title} {A note on symmetry reductions of the lindblad equation: transport in constrained open spin chains},\ }\href {https://doi.org/10.1088/1367-2630/14/7/073007} {\bibfield  {journal} {\bibinfo  {journal} {New J. of Phys.}\ }\textbf {\bibinfo {volume} {14}},\ \bibinfo {pages} {073007} (\bibinfo {year} {2012})}\BibitemShut {NoStop}%
\bibitem [{\citenamefont {Albert}\ and\ \citenamefont {Jiang}(2014)}]{Albert_2014}%
  \BibitemOpen
  \bibfield  {author} {\bibinfo {author} {\bibfnamefont {V.~V.}\ \bibnamefont {Albert}}\ and\ \bibinfo {author} {\bibfnamefont {L.}~\bibnamefont {Jiang}},\ }\bibfield  {title} {\bibinfo {title} {Symmetries and conserved quantities in lindblad master equations},\ }\href {https://doi.org/10.1103/PhysRevA.89.022118} {\bibfield  {journal} {\bibinfo  {journal} {Phys. Rev. A}\ }\textbf {\bibinfo {volume} {89}},\ \bibinfo {pages} {022118} (\bibinfo {year} {2014})}\BibitemShut {NoStop}%
\bibitem [{\citenamefont {Shibata}\ and\ \citenamefont {Katsura}(2019)}]{Shibata_2019}%
  \BibitemOpen
  \bibfield  {author} {\bibinfo {author} {\bibfnamefont {N.}~\bibnamefont {Shibata}}\ and\ \bibinfo {author} {\bibfnamefont {H.}~\bibnamefont {Katsura}},\ }\bibfield  {title} {\bibinfo {title} {Dissipative quantum ising chain as a non-hermitian ashkin-teller model},\ }\href {https://doi.org/10.1103/PhysRevB.99.224432} {\bibfield  {journal} {\bibinfo  {journal} {Phys. Rev. B}\ }\textbf {\bibinfo {volume} {99}},\ \bibinfo {pages} {224432} (\bibinfo {year} {2019})}\BibitemShut {NoStop}%
\bibitem [{\citenamefont {Ashida}\ \emph {et~al.}(2024)\citenamefont {Ashida}, \citenamefont {Furukawa},\ and\ \citenamefont {Oshikawa}}]{Ashida2024}%
  \BibitemOpen
  \bibfield  {author} {\bibinfo {author} {\bibfnamefont {Y.}~\bibnamefont {Ashida}}, \bibinfo {author} {\bibfnamefont {S.}~\bibnamefont {Furukawa}},\ and\ \bibinfo {author} {\bibfnamefont {M.}~\bibnamefont {Oshikawa}},\ }\bibfield  {title} {\bibinfo {title} {System-environment entanglement phase transitions},\ }\href {https://doi.org/10.1103/PhysRevB.110.094404} {\bibfield  {journal} {\bibinfo  {journal} {Phys. Rev. B}\ }\textbf {\bibinfo {volume} {110}},\ \bibinfo {pages} {094404} (\bibinfo {year} {2024})}\BibitemShut {NoStop}%
\bibitem [{\citenamefont {Orito}\ \emph {et~al.}(2025)\citenamefont {Orito}, \citenamefont {Kuno},\ and\ \citenamefont {Ichinose}}]{Zenodo}%
  \BibitemOpen
  \bibfield  {author} {\bibinfo {author} {\bibfnamefont {T.}~\bibnamefont {Orito}}, \bibinfo {author} {\bibfnamefont {Y.}~\bibnamefont {Kuno}},\ and\ \bibinfo {author} {\bibfnamefont {I.}~\bibnamefont {Ichinose}},\ }\bibfield  {title} {\bibinfo {title} {Dataset for "strong and weak symmetries and their spontaneous symmetry breaking in mixed states emerging from the quantum ising model under multiple decoherence"},\ }\href {https://doi.org/10.5281/zenodo.14730795} {10.5281/zenodo.14730795} (\bibinfo {year} {2025})\BibitemShut {NoStop}%
\bibitem [{\citenamefont {Weinberg}\ and\ \citenamefont {Bukov}(2017)}]{qspin1}%
  \BibitemOpen
  \bibfield  {author} {\bibinfo {author} {\bibfnamefont {P.}~\bibnamefont {Weinberg}}\ and\ \bibinfo {author} {\bibfnamefont {M.}~\bibnamefont {Bukov}},\ }\bibfield  {title} {\bibinfo {title} {{QuSpin: a Python package for dynamics and exact diagonalisation of quantum many body systems part I: spin chains}},\ }\href {https://doi.org/10.21468/SciPostPhys.2.1.003} {\bibfield  {journal} {\bibinfo  {journal} {SciPost Phys.}\ }\textbf {\bibinfo {volume} {2}},\ \bibinfo {pages} {003} (\bibinfo {year} {2017})}\BibitemShut {NoStop}%
\bibitem [{\citenamefont {Weinberg}\ and\ \citenamefont {Bukov}(2019)}]{qspin2}%
  \BibitemOpen
  \bibfield  {author} {\bibinfo {author} {\bibfnamefont {P.}~\bibnamefont {Weinberg}}\ and\ \bibinfo {author} {\bibfnamefont {M.}~\bibnamefont {Bukov}},\ }\bibfield  {title} {\bibinfo {title} {{QuSpin: a Python package for dynamics and exact diagonalisation of quantum many body systems. Part II: bosons, fermions and higher spins}},\ }\href {https://doi.org/10.21468/SciPostPhys.7.2.020} {\bibfield  {journal} {\bibinfo  {journal} {SciPost Phys.}\ }\textbf {\bibinfo {volume} {7}},\ \bibinfo {pages} {020} (\bibinfo {year} {2019})}\BibitemShut {NoStop}%
\bibitem [{\citenamefont {Calabrese}\ and\ \citenamefont {Cardy}(2004)}]{CC2004}%
  \BibitemOpen
  \bibfield  {author} {\bibinfo {author} {\bibfnamefont {P.}~\bibnamefont {Calabrese}}\ and\ \bibinfo {author} {\bibfnamefont {J.}~\bibnamefont {Cardy}},\ }\bibfield  {title} {\bibinfo {title} {Entanglement entropy and quantum field theory},\ }\href {https://doi.org/10.1088/1742-5468/2004/06/P06002} {\bibfield  {journal} {\bibinfo  {journal} {J. Stat. Mech.}\ }\textbf {\bibinfo {volume} {2004}},\ \bibinfo {pages} {P06002} (\bibinfo {year} {2004})}\BibitemShut {NoStop}%
\bibitem [{Note4()}]{Note4}%
  \BibitemOpen
  \bibinfo {note} {Technically speaking, Eq.~(\ref {CC}) is employed to estimate entanglement scaling for one-dimensional systems. Although we numerically deal with the ladder system, the quantum state can be regarded as a one-dimensional chain system from the point of view of density matrix formalism. From this viewpoint, the definition of the subsystem is consistent with Eq.~(\ref {CC}).}\BibitemShut {Stop}%
\bibitem [{\citenamefont {Vidal}\ \emph {et~al.}(2003)\citenamefont {Vidal}, \citenamefont {Latorre}, \citenamefont {Rico},\ and\ \citenamefont {Kitaev}}]{Vidal2003}%
  \BibitemOpen
  \bibfield  {author} {\bibinfo {author} {\bibfnamefont {G.}~\bibnamefont {Vidal}}, \bibinfo {author} {\bibfnamefont {J.~I.}\ \bibnamefont {Latorre}}, \bibinfo {author} {\bibfnamefont {E.}~\bibnamefont {Rico}},\ and\ \bibinfo {author} {\bibfnamefont {A.}~\bibnamefont {Kitaev}},\ }\bibfield  {title} {\bibinfo {title} {Entanglement in quantum critical phenomena},\ }\href {https://doi.org/10.1103/PhysRevLett.90.227902} {\bibfield  {journal} {\bibinfo  {journal} {Phys. Rev. Lett.}\ }\textbf {\bibinfo {volume} {90}},\ \bibinfo {pages} {227902} (\bibinfo {year} {2003})}\BibitemShut {NoStop}%
\end{thebibliography}%

\end{document}